\documentclass[12pt]{article}
\usepackage{epsfig}
\newcommand{\abstracts}[1]{{
\centering{\begin{minipage}{12.5truecm}
\normalsize\baselineskip=15pt
\centerline{\footnotesize ABSTRACT}\vspace*{0.3cm}
\parindent=20pt #1
\end{minipage}}\par}}

\newcommand{\la}{\langle}
\newcommand{\ra}{\rangle}
\newcommand{\cC}{{\cal C}}

\newcommand{\Z}{{Z \!\!\! Z}}
\newcommand{\beqn}{\begin{eqnarray}}
\newcommand{\eeqn}{\end{eqnarray}}
\newcommand{\eq}[1]{(\ref{#1})}
\newcommand{\dd}{\mbox{d}}

\newcommand{\next}{n_{\mathrm{ext}}}
\newcommand{\nint}{n_{\mathrm{int}}}

\begin{document}

~
\vspace{-1cm}
\begin{flushright}
{\large
Leipzig NTZ 09/2001\\
LU-ITP 2001/015\\
UNITU--THEP--17/2001
\\}
\vspace{0.3cm}
{\sl June 26, 2001}
\end{flushright}

\begin{center}

{\baselineskip=16pt
{\Large \bf Monopoles, confinement and deconfinement of $(2+1) D$ compact
            lattice QED in external fields}\\
 
\vspace{1cm}

{\large
M.~N.~Chernodub\footnote{maxim@heron.itep.ru}$^{\!,a}$,
E.-M.~Ilgenfritz\footnote{ilgenfri@alpha1.tphys.physik.uni-tuebingen.de}$^{\!,b}$
and A.~Schiller\footnote{Arwed.Schiller@itp.uni-leipzig.de}$^{\!,c}$}\\

\vspace{.5cm}
{ \it

$^a$ ITEP, B.Cheremushkinskaya 25, Moscow, 117259, Russia

\vspace{0.3cm}

$^b$ Institut f\"ur Theoretische Physik, Universit\"at T\"ubingen,
D-72076 T\"ubingen, Germany

\vspace{0.3cm}

$^c$ Institut f\"ur Theoretische Physik and NTZ, 
Universit\"at  Leipzig, \\ D-04109 Leipzig, Germany

}}
\end{center}
\vspace{5mm}

\abstracts{
The compact Abelian model in three space--time dimensions is
studied in the presence of external electromagnetic fields at finite
temperatures. We show that the deconfinement phase transition is
independent on the strength of the external fields. This result is in
agreement with our observation that the external fields create
small--size magnetic dipoles from the vacuum which do not influence 
the confining properties of the model. Contrary to the deconfinement phase,
the internal field in the direction of the applied external field is
attenuated in the confinement phase, this screening becomes stronger with
decreasing temperature.
}

\baselineskip=14pt
\setcounter{footnote}{0}
\renewcommand{\thefootnote}{\alph{footnote}}

\newpage 
\section{Introduction}

Compact Abelian gauge theory in three Euclidean dimensions is proven to
possess the property of permanent confinement~\cite{Polyakov,Goepfert} due
to presence of Abelian monopoles: any pair of test electric charge and
anti--charge are confined by a linear potential. The monopoles are
topological defects which appear due to the compactness of the gauge
group. In three dimensions the monopoles are instanton--like.

The confining property of the model is lost at sufficiently high
temperature. The confinement--deconfinement phase transition --- which is
expected~\cite{Sve,Coddington,Parga} to be of Koster\-litz--Thouless
type~\cite{KT} --- was studied on the lattice both
analytically~\cite{Parga} and numerically~\cite{Coddington}.  A thorough
numerical analysis of the phase transition was done in
Ref.~\cite{CIS2001a} where 
it has been demonstrated that the monopoles are
sensitive to the transition. In the confinement phase the monopoles are
observed in the plasma state while in the deconfinement phase the
monopoles appear in the form of a dilute gas of magnetic dipoles.
Similarly to the monopole plasma the dipole vacuum, although not
confining, still has a non--perturbative nature~\cite{QEDdipoles}. In the
confinement phase both monopole density and string tension differ from
semiclassical estimates~\cite{Polyakov} which neglect monopole binding.
However, an analysis of monopole clusters shows that the relation between
the string tension and the density of monopoles in {\it magnetically
charged} clusters is in reasonable agreement with those predictions.

An alternative explanation of the deconfinement phase transition,
based on Sve\-tit\-sky--Yaffe universality arguments~\cite{Sve}, was given in
Refs.~\cite{Coddington,Parga}. The phase transition was demonstrated to be
accompanied by restructuring of the $U(1)$ vortex system in an effective $2D$
Abelian spin model. In the confinement (low temperature) phase the vortices
exist in the plasma state while in the deconfinement (high temperature) phase
the vortices and anti--vortices form bound states. Both
monopole~\cite{AgasianZarembo} and vortex~\cite{Alex} binding mechanisms have
been studied in the context of the Georgi--Glashow model, 
too, a limiting case of which is the compact Abelian gauge model.

In this paper we continue to consider the deconfining mechanism by monopole
pairing  which seems to have interesting counterparts in realistic gauge
theories. The formation of  monopole pairs is qualitatively similar to the
binding of instantons in instanton  molecules with increasing temperature in
QCD suggested to be responsible for chiral symmetry
restoration~\cite{Shuryak}. An external magnetic field affects the phase
diagram of the non-Abelian theory as was shown both
analytically~\cite{QCD-external} and numerically~\cite{Cea}. In the
electroweak theory, the formation of Nambu monopole---anti--monopole pairs, a
remnant from a dense medium of disordered $Z$--vortices and Nambu monopoles
which characterizes the high temperature phase, is accompanying the transition
towards the low temperature phase~\cite{EW}. The  effects of the external
fields on the phase transition temperature and the electroweak  baryogenesis
in the electroweak theory were discussed in Ref.~\cite{EW-external}.

In three dimensional gauge theories the inclusion of external fields allows to
study the vacuum energy density in the $SU(2)$ gauge theory~\cite{Cea:1993zp}, 
vortex dynamics in the Abelian Higgs model~\cite{Dimopoulos:2001yv} and
dynamically generated fermion mass in the Abelian gauge theory with
fermionic~\cite{QED-fermions} fields. Here we study the influence of
the external electromagnetic field on the confining and monopole properties of
the compact Abelian gauge model in three dimensions.

The plan of the paper is as follows. In Section~\ref{sec:model} we describe 
the basics of the lattice formulation of the compact $U(1)$ model in $2+1$
dimensions  and how the external field is implemented. The screening of the
external electric and magnetic fields and the properties of the Polyakov loop
correlators in the presence of those fields  are studies in
Section~\ref{sec:Polloop}. Differences and similarities between magnetic
and electric fields are stressed. The string tension in the presence of the
external fields  with a flux $\next$ is investigated in
Section~\ref{sec:string} and the phase diagrams in the $\next - \beta$ plane
are presented in Section~\ref{sec:phase}, separately for magnetic and electric
fields. In the case of an applied electric field we see that the so--called
``bulk Polyakov loop'' cannot play any longer the usual role in
localizing the deconfining transition. Results for various monopole properties
are collected in Section~\ref{sec:monopole:properties}. Our conclusions are
summarized in the last Section.


\section{The compact lattice model in external fields} 
\label{sec:model}

We study $2+1$ dimensional compact lattice electrodynamics in constant
external electromagnetic fields. Following  Ref.~\cite{DamgaardHeller} we use
the action  
\beqn
  S[\theta, \theta^{\mathrm{ext}}] = - \beta \sum\limits_p \cos\Bigl(\theta_p -
  \theta^{\mathrm{ext}}_p\Bigr)\,,
  \label{eq:extint}
\eeqn
where $\theta_p$ is the $U(1)$ field strength tensor represented by the curl
of the compact link field $\theta_l$, and $\theta^{\mathrm{ext}}_p$ is the
field strength corresponding to the external field. $\beta$ is the lattice
coupling constant related to the lattice  spacing $a$ and the continuum
coupling constant $g_3$ of the $3D$ theory as follows: 
\beqn
 \beta = \frac{1}{a\, g^2_3}\,.
\label{beta}
\eeqn
  
At finite temperature the lattice is asymmetric, $L^2_s\times L_t$, $L_t <
L_s$; $L_1=L_2=L_s$ and $L_3=L_t$ are the spatial and  temporal extensions of
the lattice, respectively. In the limit $L_s \to \infty$ the temporal
extension of the lattice $L_t$ is related to the physical temperature, $L_t =
1 \slash (T a)$. Using eq.~\eq{beta} the temperature is given via the lattice
parameters as follows: 
\beqn
  \frac{T}{g^2_3} = \frac{\beta}{L_t}\,.
  \label{temp}
\eeqn

Note that in $2+1$ dimensions there is no symmetry anymore between the three
components of the field strength tensor. The closest relative of the true
magnetic field is $F_{12}$ distinct from the others, while there is still a
symmetry between $F_{13}$ and $F_{23}$. With this distinction in mind one can
conditionally call them the ``magnetic'' and ``electric'' components of the
field strength tensor, respectively. As long as one does not introduce
external fields, even at finite temperature there was no need to distinguish
between them.  

Without loss of generality we represent an external electric field $E$ as a
field directly coupled via (\ref{eq:extint}) to plaquettes lying in the $31$
plane. The value of the field is quantized~\cite{DamgaardHeller}:  
\beqn
  E = 
  \theta^{\mathrm{ext}}_{31}= \frac{2\pi n_E}{L_1\, L_3}\,,\quad 
  n_E \in \Z\,,\quad
  \theta^{\mathrm{ext}}_{12}= 
  \theta^{\mathrm{ext}}_{23}=0\,.
  \label{theta:quantE}
\eeqn
To study the influence of a magnetic field (and to draw parallels to 
analogous studies in $4D$~\cite{Bari}) we implement a magnetic field $B$ 
as a field directly coupled via (\ref{eq:extint}) to plaquettes lying 
in the $12$ plane,
\beqn
  B=
   \theta^{\mathrm{ext}}_{12}= 
  \frac{2\pi n_M }{L_1\, L_2}\,,\quad 
   n_M \in \Z\,,\quad
  \theta^{\mathrm{ext}}_{31}= 
  \theta^{\mathrm{ext}}_{23}=0\,.
  \label{theta:quantM}
\eeqn

The quantization of the external fields (\ref{theta:quantE}),
(\ref{theta:quantM}) is necessary to match with the periodic boundary
conditions imposed on the lattice~\cite{DamgaardHeller}. Indeed, these
conditions imply that the rectangular Wilson loop of the size $L_i \times L_j$
in the $ij$ plane must be equal to unity. On the other hand the Wilson loop is
equal to  $\exp\{i \Phi\}$, where $\Phi$ is the flux going through the $ij$
plane. Therefore, we get the quantization condition for the flux, $\Phi = 2
\pi \next$, $\next \in \Z$, which implies in turn the relations
(\ref{theta:quantE}), (\ref{theta:quantM}) for the constant electric or
magnetic field, respectively. We use the notation $n_{E/M}$ instead of
$\next$ where we want to emphasize the electric/magnetic nature of
the external field. 

Naturally, from the form of the action~\eq{eq:extint} the number of external
fluxes is restricted, $0< \next < L_i L_j \slash 2$. In our analysis we
restrict that number to $0< \next < L_i L_j \slash 4$. One should note
however, that the largest considered fluxes correspond to  lattice artifacts.
Indeed, the physical electromagnetic field must have a strength smaller than
the scale corresponding to the lattice ultraviolet cutoff, $a^{-2}$. This
leads to the restriction of the lattice field strength: $\theta_{\mathrm{ext}}
\ll 1$, or, $\next \ll n_{\mathrm{max}} = [L_i L_j \slash (2\pi)]$, to avoid
too large external fields. 

As in the case of zero field~\cite{CIS2001a} we restrict ourselves to a finite
temperature lattice $32^2 \times 8$ varying the strength of the constant
external electric or magnetic field and considering the lattice gauge coupling
range $0.1 \leq \beta \leq 3$. For this particular lattice size we get the
upper bounds for the electric and magnetic fields: $n_E^{\mathrm{max}} = 40$
and $n_M^{\mathrm{max}}= 163$, respectively. From our studies at zero external
fields we~\cite{CIS2001a} found the deconfinement phase transition at $\beta_c
= 2.346(2)$ using the Polyakov loop susceptibility in satisfactory agreement
with Ref.~\cite{Coddington}.

We briefly recall the Monte Carlo algorithm used and described already in
Ref.~\cite{CIS2001a}. The algorithm combines a local Monte Carlo step with a
global refreshment step to improve ergodicity. The local Monte Carlo
algorithm is based on a 5--hit Metropolis update sweep followed by a
microcanonical sweep. Following  Ref.~\cite{DamgaardHeller}, the global
refreshment step consists of an attempt to add an additional unit of flux 
of randomly chosen direction and sign ({\it i.e.} adding a gauge field
$\tilde{\theta}_i$ constructed over the  whole lattice to the previous one, 
$\theta_i \rightarrow {[\theta_i + \tilde{\theta}_i]}_{{\mathrm{mod}} 2\pi} $,
subject to a global Metropolis acceptance check. The acceptance rate for a
global update does not depend on the external flux $\next$ and it is roughly
equal  for global shifts of the electric or magnetic flux. However, it changes
with  $\beta$, reducing from $0.58 \dots 0.6$ at $\beta=1.0$ to $0.18 \dots
0.20$ at $\beta=2.9$. 


\section{Polyakov loop correlators and screening}
\label{sec:Polloop}

Usually, the Polyakov loop 
\beqn
   L(\mathbf x) = \exp \Bigl\{ i \int\limits^{L_t}_0 \dd t \, 
                  A_0 (\mathbf x, t)\Bigr\} 
\eeqn
is used as a the basic quantity to probe the confining properties. The
integration runs along a loop $l_{\vec x}$ parallel to the time axis and
located at the spatial $2D$ coordinate $\mathbf x$. The Polyakov loop
operator inserts an infinitely heavy test particle with unit electric charge
into the vacuum of the theory. The $v.e.v.$ of this operator expresses the
free energy $F$ of the inserted particle, $\la L(\mathbf R) \ra = \exp(- F
\slash T)$ and is an order parameter to signal deconfinement.

In the absence of an external field the correlation function between two
Polyakov loops can be expressed via the interaction potential $V(R)$ between
infinitely heavy electric charge and anti--charge: 
\beqn
  \la L(\mathbf 0) L^{*}(\mathbf R)\ra \propto
  \exp\{ - L_t \, V(\mathbf R)\}\,.
  \label{ll1}
\eeqn
The leading behavior of the potential $V(\mathbf R)$ in the low temperature
phase corresponds to a linearly rising potential, $V(R) = \sigma \, R$, where
$\sigma$ is the tension of the string between the test electric charges
separated by the distance $R = |\mathbf{R}|$. In the high temperature phase
the potential is of Coulomb type (it rises logarithmically with increasing
distance between test particles).

Now we discuss the Polyakov loop correlations in both electric and magnetic
external fields. We introduce a $3D$ vector combining electric and magnetic
components by  
\beqn
  \vec F =(F_{23},F_{31},F_{12}) = (E_x,E_y,B_z)\,.
  \label{vecF}
\eeqn
As we will immediately see, an external (constant) electric field 
substantially modifies the behavior of the  Polyakov loop correlator~\eq{ll1}
already {\it on the tree level}. 

Indeed, the electromagnetic field inside the medium can be decomposed into
two parts,  $\vec F^{\mathrm{int}} + \vec F^{q}$.
Here $\vec F^{\mathrm{int}}$ is the mean field inside the medium, which is
non--zero due to the presence of the external electromagnetic field. The field
$\vec F^{q}$ corresponds to the quantum fluctuations around the mean value
$\vec F^{\mathrm{int}}$. Due to the Abelian nature of the Polyakov loop the
correlation function~\eq{ll1} can be written as product of two contributions 
\beqn
  {\la L(\mathbf 0) L^*(\mathbf R)\ra}_{{\vec F^{\mathrm{ext}}}} & = & 
  {\la \exp \{i \oint\nolimits_\cC
  \dd x_\mu A_\mu (x)\} \ra }_{{\vec F^{\mathrm{ext}}}} = 
  {\mathrm e}^{i \Phi_\cC (\vec F^{\mathrm{int}} )} \cdot
  G(\mathbf R, {\vec F^{\mathrm{ext}}})\,, 
  \label{llH1} 
  \\
  G(\mathbf R, {\vec F^{\mathrm{ext}}} ) & = & {\la L(\mathbf 0) L^*(\mathbf
  R)  \ra}^{q}_{{\vec F^{\mathrm{ext}}}} = 
  {\la {\mathrm e}^{i \Phi_\cC(\vec F^{q})} \ra}_{{\vec
  F^{\mathrm{ext}}}}^{q}\,,  
  \label{llH2}
\eeqn
where $\cC = l_{\mathbf 0} - l_{\mathbf R}$ is the contour corresponding to the
test  particle trajectories, and 
\beqn
  \Phi_\cC(\vec F) = \oint\nolimits_\cC \dd x_\mu A_\mu (x) = 
  \int\nolimits_{\Sigma_\cC} \dd \sigma_\mu \, F_\mu (x)\,,
  \label{PhiC}
\eeqn
is the flux of the electromagnetic field $F$ which goes through the surface 
$\Sigma_\cC$ spanned on the contour $\cC$. The subscript ${\vec
F^{\mathrm{ext}}}$ in eq.~\eq{llH2} indicates,  that the vacuum expectation
value of the quantum part of the Polyakov loop has been taken with the 
action eq.~\eq{eq:extint} corresponding to non--zero external flux.

The correlator~\eq{llH2} is taken over quantum fluctuations only, as
indicated by the superscript $q$.  Note that by definition the
field $A^{q}$ fluctuates near the minimum of the action. Thus negative modes
are absent and the potential $V(\mathbf R; {\vec F^{\mathrm{ext}}}) \propto 
- \log G(\mathbf R , {\vec F^{\mathrm{ext}}}) / L_t$ must be real. 
The dynamics of the monopoles affects the quantum fluctuations and
consequently the potential $V$. The external fields disturb the monopole
medium  and therefore we may expect that in general the inter--particle
potential $V$ may get influenced by the external fluxes.

Summarizing, the quantum average of the Polyakov loop in the presence of the
external constant electromagnetic field is given by the following formula:
\beqn
  {\la L(\mathbf 0) L^{*}(\mathbf R)\ra }_{\vec F^{\mathrm{ext}}} 
  \propto  
  \exp \Bigl\{i 
     \Phi_\cC(\vec F^{\mathrm{int}}) 
  - L_t \, V(\mathbf R; {\vec F^{\mathrm{ext}}} )
  \Bigr\}\,,
  \label{ll:H}
\eeqn
where a {\it non--vanishing} flux $\Phi_\cC(\vec F^{\mathrm{int}})$ 
pierces the surface $\Sigma_\cC$ spanned  by the contour $\cC$ and the
trajectories of the test particles are placed along the (temporal)
$z$--direction.

The properties of the internal fields inside the system are important to
understand the behavior of the Polyakov loop correlators. For external fields
given by eqs.~(\ref{theta:quantE}), (\ref{theta:quantM}) the total 
external fluxes through the corresponding planes are quantized.
Analogously, the internal electric or magnetic fluxes through any plane in
any gauge field configuration are quantized as well (the considerations
similar to those in Section~\ref{sec:model} are applicable in this case as
well). Note that the global update step is consistent with the internal flux
quantization since the update allows to change the total flux of the system by
just one unit. 

{}From now on we study only two possibilities for the external field,
an electric field in $y$ direction and magnetic field in $z$ direction,
\beqn
  \vec F^{\mathrm{ext}}=(0,E,0)\,,\quad \mbox{or} 
  \quad \vec F^{\mathrm{ext}}=(0,0,B)   
\eeqn
It is clear that only the electric component contributes to the 
flux $\Phi_\cC(\vec F^{\mathrm{int}})$ in eq.~\eq{ll:H}. 
In our case the mean internal electric and magnetic fields
which are the actual average fields present in the medium
 are given by 
$\la \theta_{31} \ra$ and $\la \theta_{12} \ra$, respectively, where
\beqn
  \la \theta_{ik}\ra=
   \la \ \frac{1}{L_s^2L_t}
   \sum_{\vec x}[\theta_{\vec x ,ik}]_{{\mathrm{mod}} 2\pi} \ \ra\,.
   \label{eq:intfield}
\eeqn

\begin{figure*}[!htb]
  \begin{center}
    \begin{tabular}{cc}
      \epsfxsize=7.0cm \epsffile{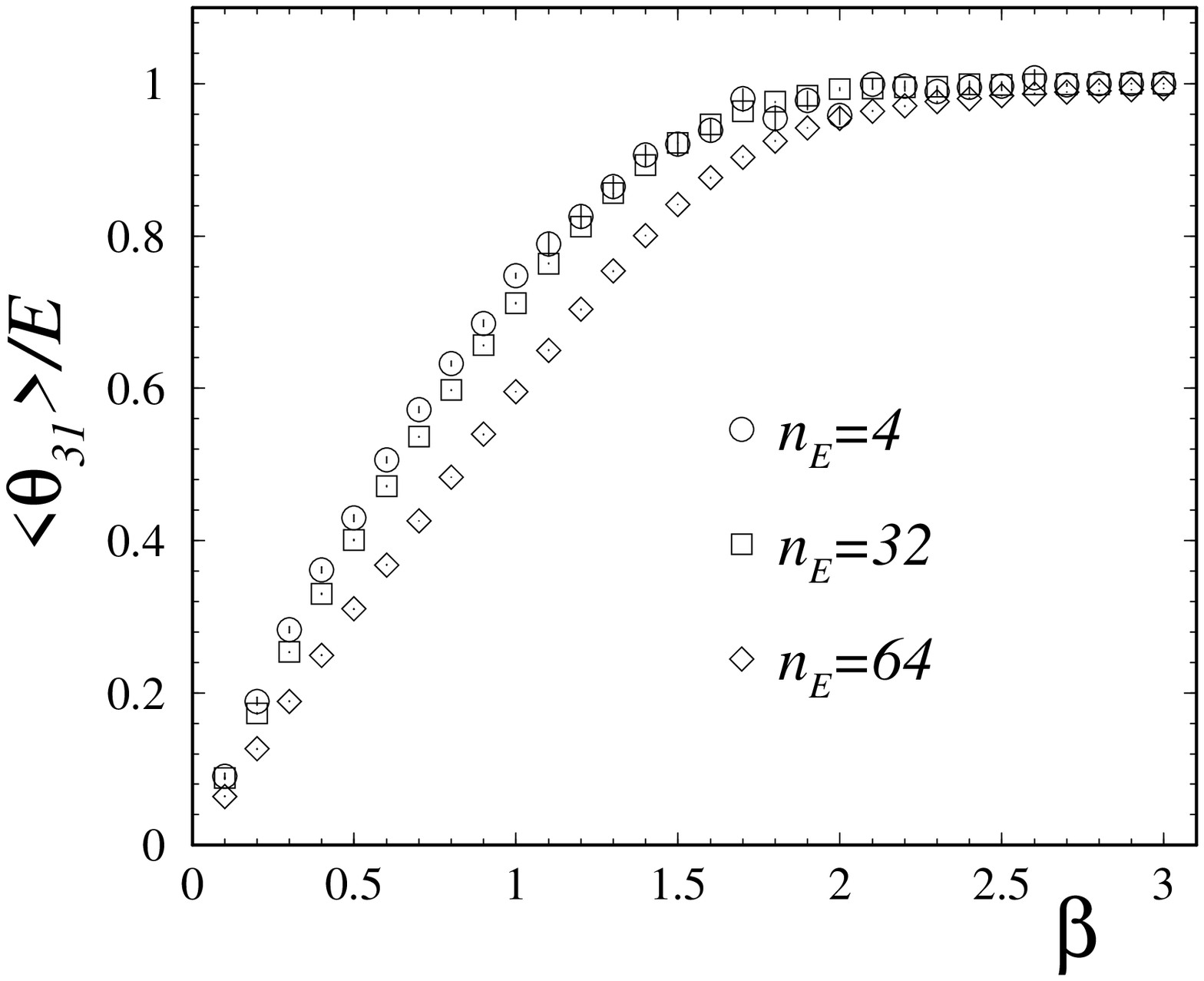} & 
      \epsfxsize=7.0cm \epsffile{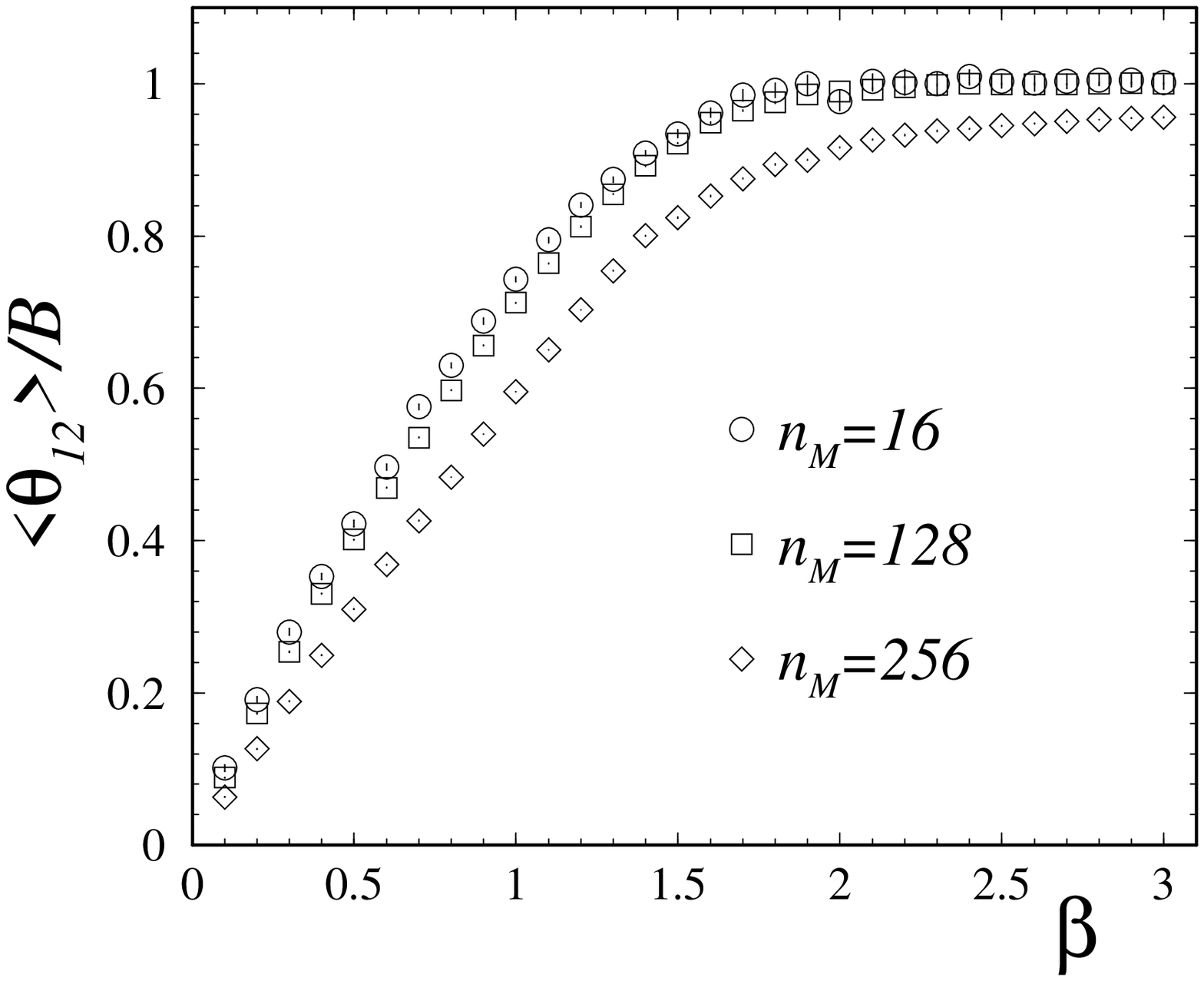} \\
      (a) & \hspace{1.5cm}  (b) \\
    \end{tabular}
  \end{center}
  \vspace{-0.5cm}
  \caption{The normalized internal (a) electric and (b) magnetic fields
         {\it vs.} $\beta$ at fixed 
         $\next$.}
  \label{fig:theta}
\end{figure*}

Following Ref.~\cite{Cea:1991ag} we show in Figure~\ref{fig:theta}
the normalized average internal electromagnetic fields
$\la \theta_{31} \ra \slash E = F_2^{\mathrm{int}} \slash E$ 
and $\la \theta_{12} \ra \slash B = F_3^{\mathrm{int}} \slash B$ 
along the directions of the applied fields as function of the gauge
coupling $\beta$ for various external field fluxes $\next$.
One can see that in the confinement phase the internal fields are 
much weaker than the external fields while in the deconfinement phase the
internal and external fields almost coincide with each other.  
The measured average internal fields in the directions transverse to the
applied external field remain zero.

The attenuation of the external fields in the confinement phase resulting
in weaker average fields in the medium happens due to the monopole
plasma~\cite{Polyakov} and can be called screening: the monopoles produce
a finite correlation length $\xi$. If we would apply an external field to
the box with the monopole plasma using frozen boundary conditions then the
field would be greatly suppressed inside the media at distances (from the
box boundaries) larger than $\xi$. Thus the averaged field inside the box
with the plasma should be smaller than the external field. The smaller the
correlation length the smaller the averaged field inside the media should
be.

As the temperature increases the monopoles form more and more magnetic dipole
bound states. In the dipole plasma the screening is absent~\cite{no-screening}
and the correlation length is infinite. The internal field is expected to be
equal to the external one what is clearly seen in Figures~\ref{fig:theta} for
external fields of small strength. Moreover, the screening of the
external magnetic and electric field with the same strength (related to each
other as $n_E = n_M \slash 4$ due to difference in the lattice extensions in
spatial and temporal directions) is very similar for weak external fields.
This happens because the nature of the monopoles is not purely ``magnetic'' at
finite temperature. At large magnetic fields the response of the media is
different in magnetic and electric cases. In
Section~\ref{sec:monopole:properties} we show that the observed features of
the screening are qualitatively related to specific properties of the
monopoles and their bound states.

Let us consider the properties of the correlator~\eq{ll:H} on a lattice 
of finite size for a non--vanishing external electric field $E$
using the lattice Polyakov loop at position ${\mathbf{x}}=(x,y)$
\beqn
   L(\mathbf x)= \exp{\left( i  \sum_{z=1}^{L_t} \theta_3( \mathbf x ,z)
   \right)} \,. 
\eeqn
We put the test charge and anti--charge at the points $(0,0)$ and $(x,y)$,
respectively. The fluxes $\Phi_\cC(\theta_{31})$ going through the surface
$\Sigma_\cC$ spanned  between the test particle trajectories depend on the
internal electric field $\theta_{31}$, they are quantized and peaked around an
average value. Therefore, they can adequately be described by taking into
account only the ``most probable'' flux state labeled by the integer  
$\nint=\nint(\next,\beta,L_i)$ which depends on the strength of the external
field, temperature and lattice geometry. 
In that case  $\Phi_\cC = 2 \pi \, \nint \, x \, \slash L_s$ and
$E_{\mathrm{int}}=\la \theta_{31}\ra=2 \pi \nint \, \slash (L_s\,L_t)$.  
The Polyakov loop correlator reads as follows,   
\beqn 
  {\la L(0,0) \, L^{*}(x,y)\ra }_E  \propto
   \exp\Bigl\{2 \pi i \nint \,  \frac{x}{L_s}  - L_t \,
  V(x,y;E)\Bigr\}\,, 
\label{ll:H:lat}
\eeqn
with the oscillating part of the correlator defined by the electric component 
of the internal field.

Eq.~(\ref{ll:H:lat}) simultaneously characterizes both the screening of the
external field (phase factor) and the potential of the test electric charges
(modulus). We expect that in the deconfinement phase the external field is
not screened and the correlator is dominated by  $\nint\approx n_E$.
In the confinement phase the field must be screened and $\nint$ becomes
smaller with decreasing $\beta$ (temperature).

Thus the Polyakov correlator in the deconfinement phase is an oscillating
sign--chan\-ging quantity at fixed $y$ coordinate as function of the $x$
coordinate. The maxima of the oscillations become damped with increasing
$x$ due to the Coulombic interaction between the test particles.

\begin{figure}[!htb]
  \begin{center}
    \begin{tabular}{cc}
    \epsfxsize=7.5cm \epsffile{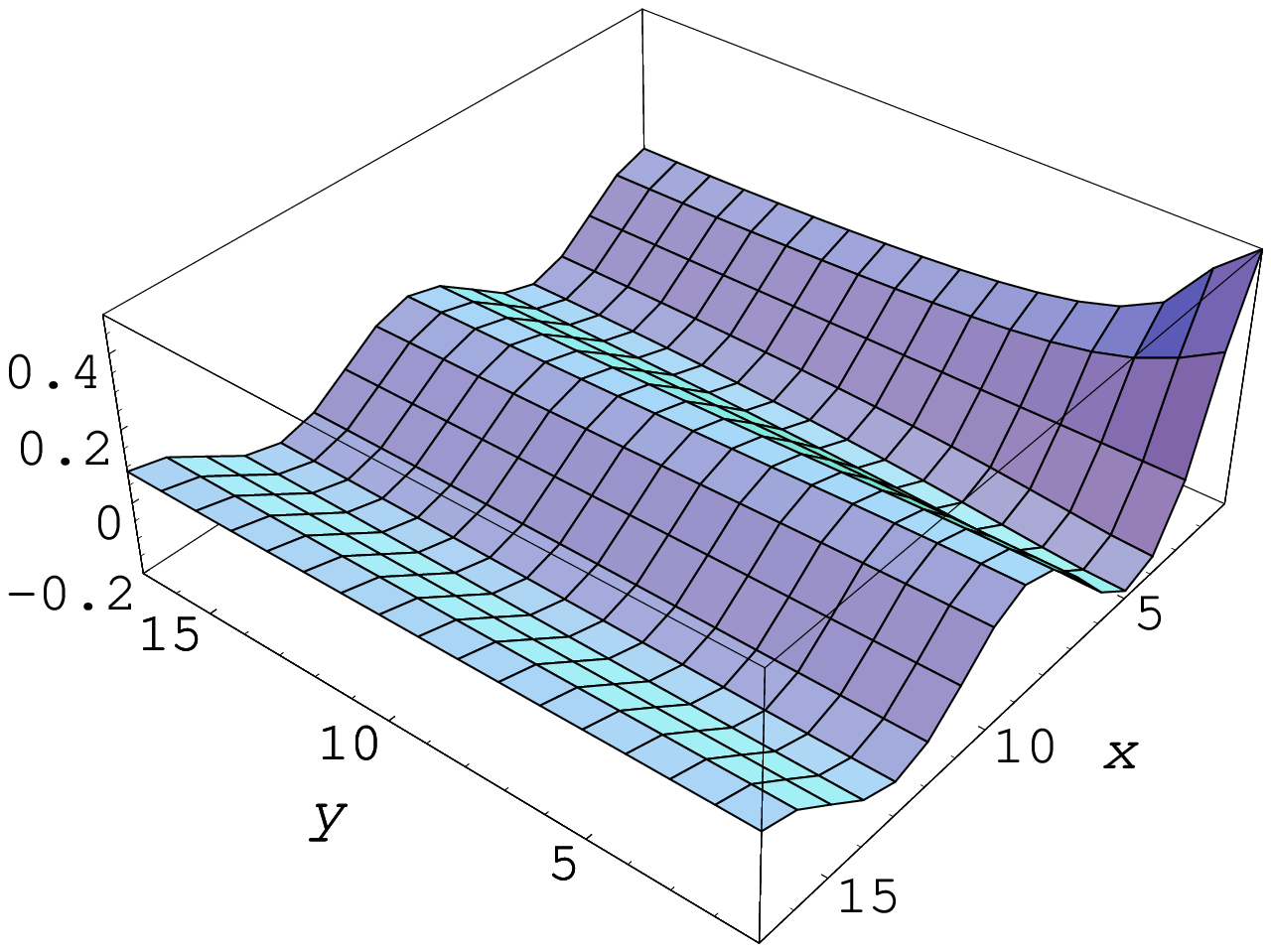} &
    \epsfxsize=7.5cm \epsffile{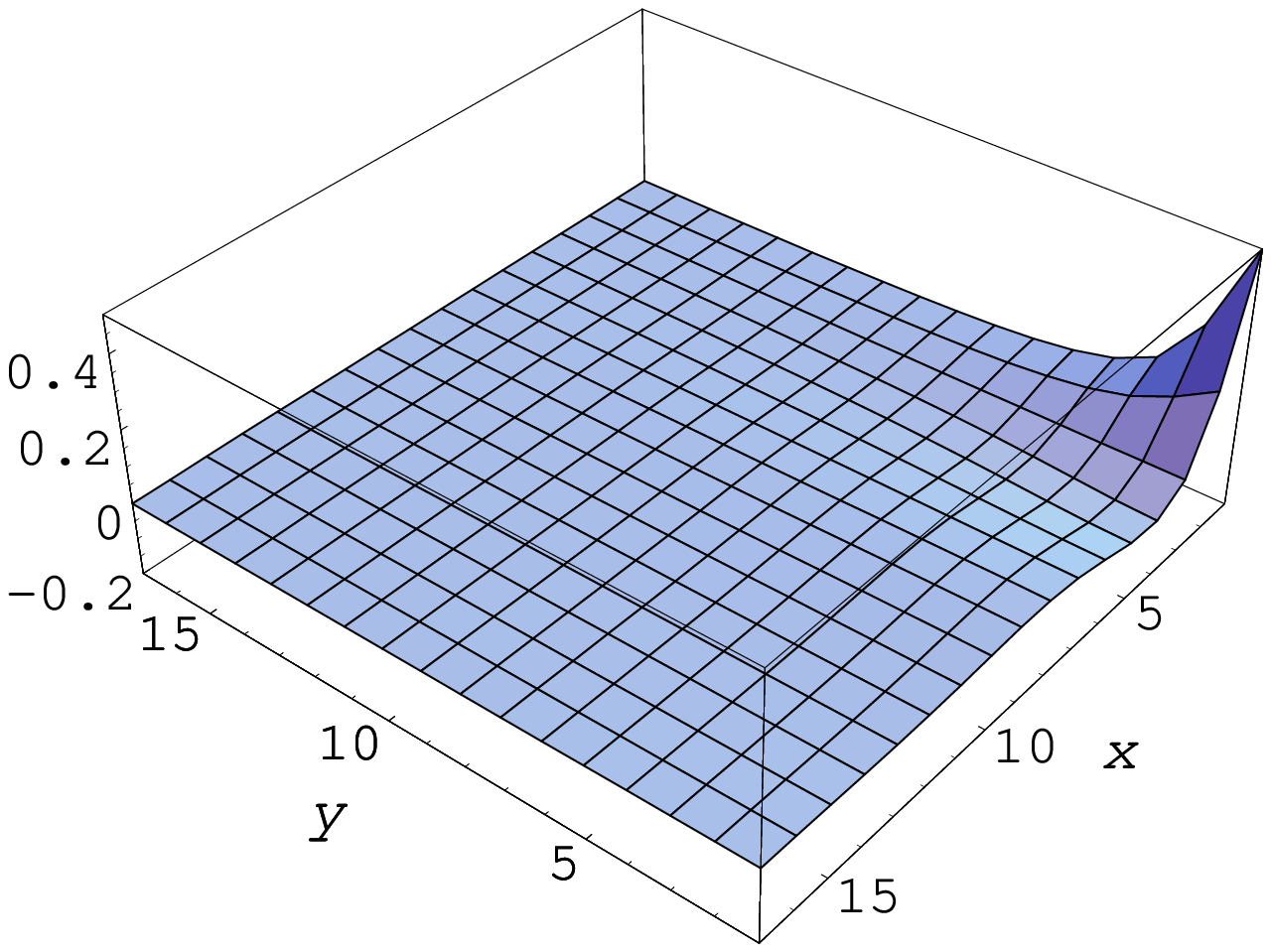} \\
    (a) & \hspace{1.5cm}  (b) \\
    \end{tabular}
  \end{center}
  \vspace{-0.5cm} 
  \caption{Real part of the Polyakov loop correlator in the $x-y$
  plane~(\ref{ll:H:lat}) for an external electric flux
  $n_E=4$ in (a) deconfinement, $\beta=2.6$ and (b)
  confinement, $\beta=2.0$, phases. Half of the lattice $32^2 \times 8$ is
  shown.}   
\label{fig:polloop-xy} 
\end{figure}

The correlations at fixed $x$ and changing $y$ should be of the same sign and 
decay with increasing $y$. This behavior is nicely illustrated by the (real
part of) the Polyakov loop correlators~\eq{ll:H:lat} in the $x - y$ plane
shown in Figure~\ref{fig:polloop-xy}(a)
for the deconfinement phase. The measurement was performed in an external
electric field with flux $n_E= 4$.

In the confinement phase we expect that the sign fluctuations of the
correlator should be strongly suppressed due to area--law decay. 
Thus the correlator must go to zero rapidly with increasing distance between
charge and anti--charge, $\sqrt{x^2 + y^2}$. 
The corresponding correlator, presented in Figure~\ref{fig:polloop-xy}(b) at
$\beta=2.0$, shows the expected suppression.  
In addition, we observe a small local maximum at $x \approx 8$ and $y=0$
arising from a flux of value  $\nint$ close to 4 in agreement with
the behavior of the internal fields in Figure~\ref{fig:theta}
for this relatively large $\beta$ value.

Finally, let us comment here that an induced internal magnetic field
(parallel to its parental external field) which is directed {\it along} the
Polyakov loop cannot  contribute to the ``classical'' phase $\Phi_\cC$ and,
therefore, oscillations cannot appear. Thus, in an external
magnetic field, the correlator is rapidly approaching zero with
increasing spatial distance.

\section{String tension in the presence of fields}
\label{sec:string}

According to Ref.~\cite{Valya} the point--point correlation of the two
Polyakov loops in the absence of an external field can be  parametrized as
follows\footnote{Here and below we take into account only the lowest ``mass
state'' corresponding to the actual string tension
tension, unless specified otherwise.}: 
\beqn
  \la L(x_1,y_1) L^*(x_2,y_2) \ra = {\mathrm{const}} \cdot 
  \sum\limits_{p_{1,2}} \frac{{\mathrm e}^{i p_1 (x_1 - x_2) 
  + i p_2 (y_1 - y_2)}}{1 - \cos p_1 - \cos p_2 + \cosh (\sigma L_t)} +
  \dots\,,
  \label{mitr}
\eeqn
where the sum runs over all possible momenta, $p_{1,2} = 0, \dots, 
2 \pi (L_s-1) \slash L_s$ and $\sigma$ denotes the ``temporal'' string tension.
 
To evaluate the string tension, we use two Polyakov ``plane'' operators which
are defined as a sum over Polyakov loop positions along and perpendicular to
the external field.  As an example we take the electric field in $y$ direction.
Then
\beqn
  L_\parallel(x) = \sum\limits_{y=1}^{L_s} L(x,y) \,, \quad
  L_\perp(y) = \sum\limits_{x=1}^{L_s} L(x,y)\,.
\eeqn
The correlator of the plane--plane correlators may be written as a sum over
the corresponding positions of the point--point correlation functions,
\beqn
  \la L_\parallel(0) L^*_\parallel(x)\ra 
  & = & \sum\limits_{y_{1,2}=1}^{L_s} \la L(0,y_1) L^*(x,y_2) \ra\,, 
  \label{ll:parallel}
  \\
  \la L_\perp(0) L^*_\perp(y)\ra 
  & = & \sum\limits_{x_{1,2}=1}^{L_s} \la L(x_1,0) L^*(x_2,y) \ra\,.
  \label{ll:perp}
\eeqn
 
In the presence of an electric field $E$ along the $y$ axis this correlator
must be modified in accordance with our discussion in Section~\ref{sec:Polloop}
\beqn
  & & \la L(x_1,y_1) L^*(x_2,y_2)\ra \to {\la L(x_1,y_1)
  L^*(x_2,y_2)\ra}_E \nonumber\\
  & = &  
  \exp  {\left( - 2 \pi i \nint  \frac{x_1 - x_2}{L_s} \right)} \, 
  {\la L(x_1,y_1) L^*(x_2,y_2)\ra}^{q}_{n_E} \,.   
  \label{mitr:H}
\eeqn
The subscript $n_E$ indicates that the Polyakov loops are
calculated in the state with 
an external 
electric flux $n_E$ in the $y$ direction. 
The internal 
electric flux $n_{\mathrm{int}}$ depends implicitly on that external flux. 
Combining eqs.~(\ref{mitr}), (\ref{mitr:H}) with
eqs.~(\ref{ll:parallel}), (\ref{ll:perp})  and taking the sums over the
momenta $p_{1,2}$ explicitly, we get:   
\beqn 
  {\la L_\parallel(0) L^*_\parallel(x)\ra}_{n_E} & = & {\mathrm{const}} \cdot
  {\mathrm e}^{2 \pi i n_{\mathrm{int}}\, x \slash L_s } \, \cosh
  \Bigl[\sigma\, L_t \, \Bigl(x - \frac{L_s}{2}\Bigr)\Bigr]\,,
  \label{ll:parallel:2}
  \\ 
  {\la L_\perp(0) L^*_\perp(y)\ra}_{n_E} & = & {\mathrm{const}} \cdot 
  \cosh \Bigl[\sigma_{\mathrm{eff}}(\sigma,n_{\mathrm{int}})\, L_t \,
  \Bigl(y - \frac{L_s}{2}\Bigr)\Bigr]\,, 
  \label{ll:perp:2} 
\eeqn 
where 
formally a string tension coefficient
$\sigma_{\mathrm{eff}}(\sigma,n_{\mathrm{int}})$ for the plane--plane
correlator perpendicular to the field is:    \beqn
  \sigma_{\mathrm{eff}}(\sigma,n_\mathrm{int}) = \frac{1}{L_t} \, 
  {\mathrm{arccosh}} \Bigl[
  \cosh(\sigma\, L_t) - \cos(2 \pi n_{\mathrm{int}}\, \slash L_s) + 1 \Bigr]\,.  
  \label{eff:sigma}
\eeqn 
Thus in the presence of an external electric field the plane--plane
Polyakov loop correlator parallel to the electric field oscillates with 
a decreasing amplitude.

The plane--plane correlator perpendicular to the field decreases
exponentially (without oscillations) as function of the distance between the
planes. The decrease of this correlator is controlled by an {\it  effective}
string tension~\eq{eff:sigma} which is a function of the external electric
field (via $\nint$) and the actual string tension. The essential message here
is that this effective string tension does not tell anything about confinement
properties, since the confinement is described by the real string tension
$\sigma$. Note that the effect of the external electric field is absent if
the internal flux is equal to quantized values   
\beqn 
  \nint=N \cdot L_s\,, \quad N \in \Z\,. 
  \label{cond:N:E}
\eeqn

The origin of the $\sigma_{\mathrm{eff}}$ dependence on the external
electric field  strength is related to the oscillatory behavior of the
Polyakov loop correlator~\eq{ll:H:lat}. Indeed, the bulk correlator $\la
|L|^2 \ra_E$   (defined below in eq.~\eq{bulk:pol:loop}) comprises all
possible Polyakov loop correlators. At small and increasing $\nint$ the sum of
the  oscillating quantities leads to a suppression of $\la |L|^2 \ra_E$ 
while at $\nint$ satisfying the special condition~\eq{cond:N:E} the
oscillations disappear and as a result the  effective string
tension~\eq{eff:sigma} coincides with the string tension in the absence of 
the external electric field.    

To get rid of the direct influence of the electric field on the string 
tension~\eq{eff:sigma} we use   only the real part of the Polyakov
plane--plane correlator which is  parallel to the external field. We fit the
numerical data by the real part of  eq.~\eq{ll:parallel:2} using (besides
trivial prefactors) $\sigma$ and  $\nint \in \Z$ as fitting 
parameters. At the considered values of $\beta \ge 1.6$ we find that the
best fit is  given by $\nint=\next$ in accordance with our discussion in 
Section~\ref{sec:Polloop}. 

\begin{figure*}[!htb]
  \begin{center}
     \begin{tabular}{cc}
        \epsfxsize=7.0cm \epsffile{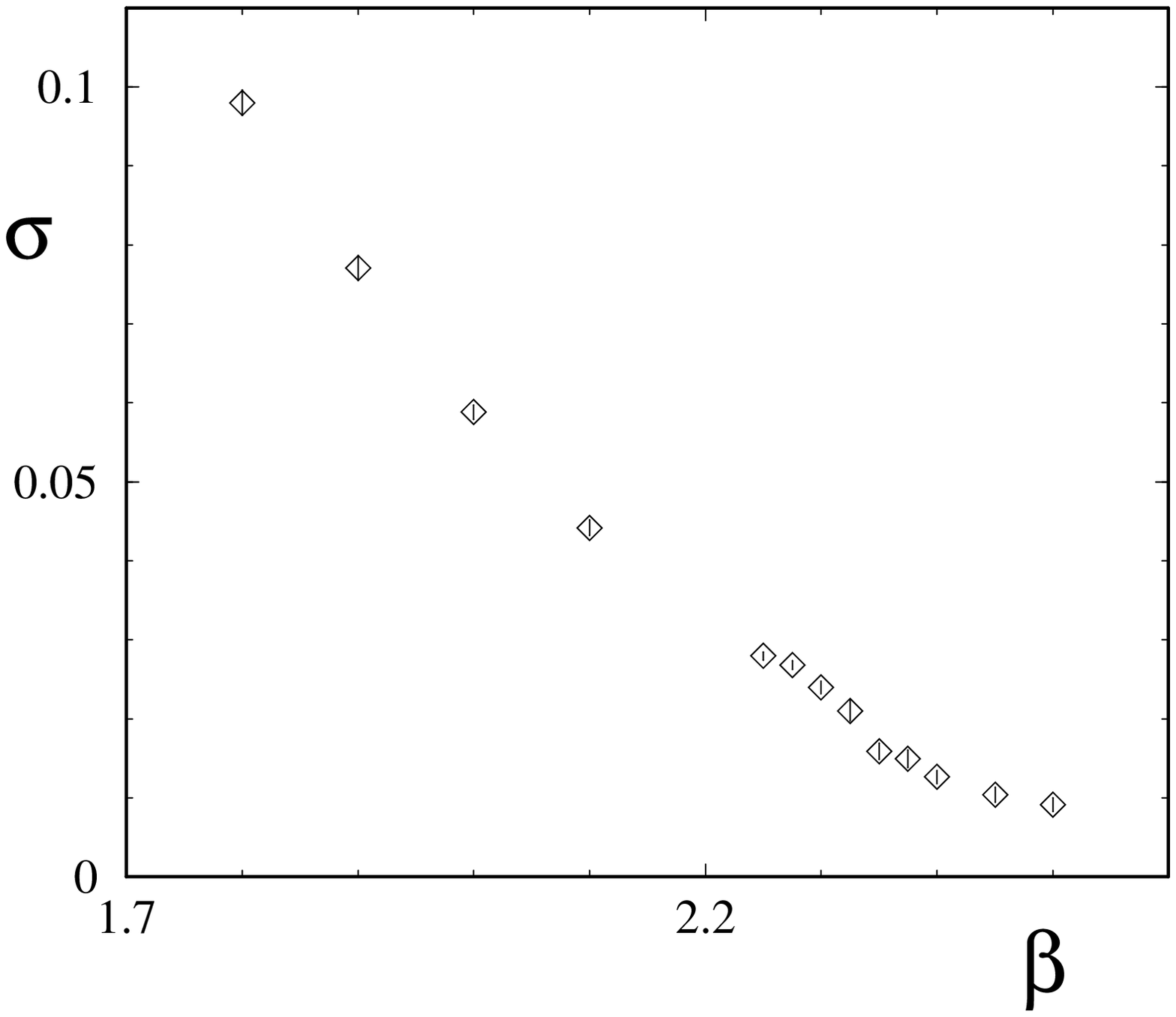}  &  
        \epsfxsize=7.0cm \epsffile{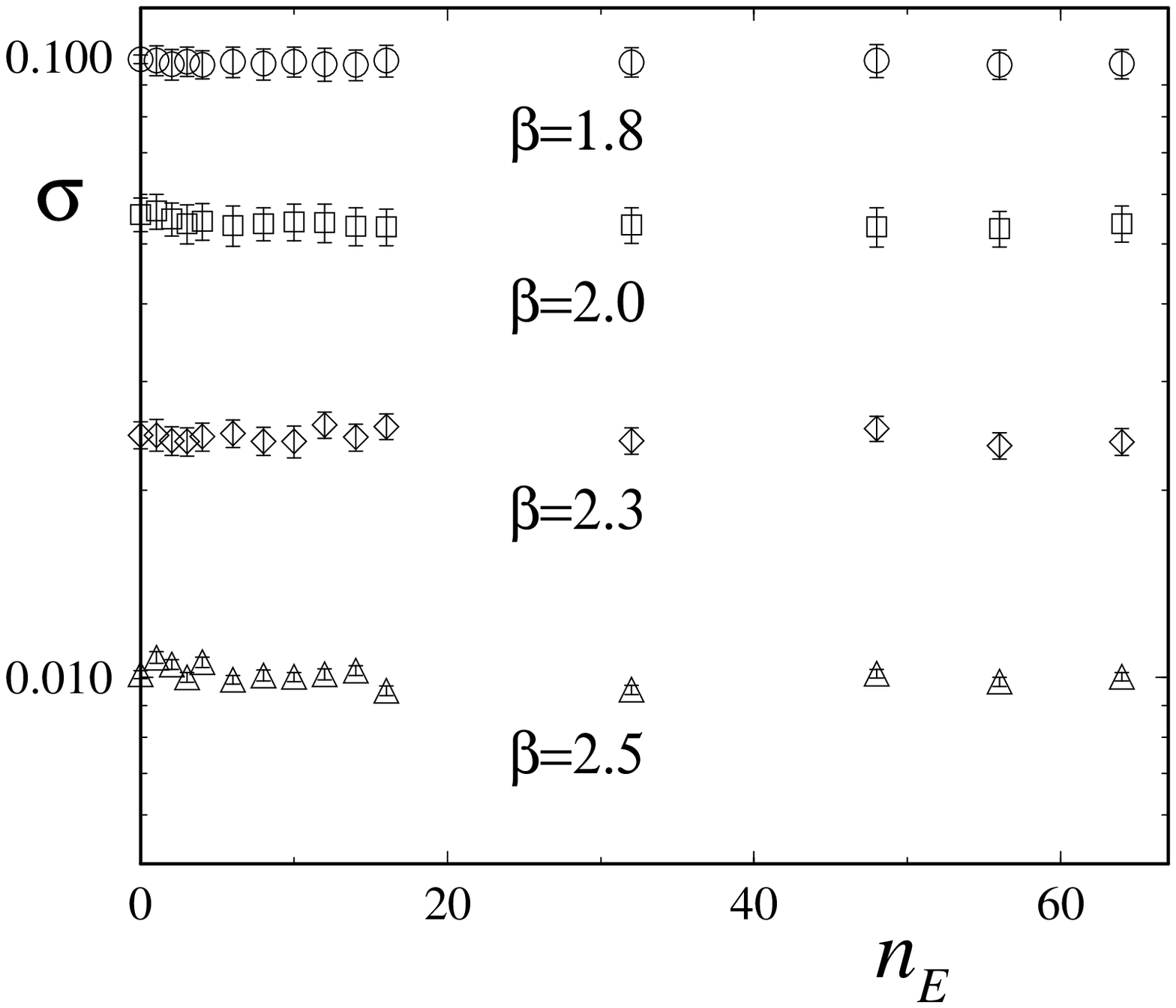} \\
            (a) & \hspace{1.5cm}  (b) \\
     \end{tabular}
   \end{center}
   \vspace{-0.5cm}
   \caption{(a) The string tension {\it vs.} $\beta$ without external field
   taken from Ref.~\cite{CIS2001a}. 
   (b) Fitted string tension for various $\beta$ values as function of the
   external electric flux $n_E$.}
   \label{fig:string:tension}
\end{figure*}

In the absence of external fields, the dependence of the string tension on
$\beta$ has been found in  Ref.~\cite{CIS2001a} and it is shown in
Figure~\ref{fig:string:tension}(a). 
The string tension is a rapidly decreasing function: it is non--zero in the
low temperature phase (low $\beta$) and becomes very small (due to finite
size effects)  in the high temperature phase. {}From
Figure~\ref{fig:string:tension}(b) we conclude that  within errors the string
tension does not depend on the external flux $n_E$. This observation allows us
to conclude that an external electric field in the compact $(2+1)D$ Abelian
gauge theory does not change the confinement behavior compared to the zero
field case. 

For an external magnetic field (in the ``temporal'' $z$ direction) the
oscillating factor in the Polyakov loop correlator is absent.
Therefore, we have used eq.~\eq{ll:parallel:2} with $n_{\mathrm{int}}=0$ to 
fit the plane--plane correlation functions of the Polyakov lines.
The numerical results for the string tension $\sigma$ are similar to those in
the case of the electric field shown in Figure~\ref{fig:string:tension}(b): 
within errors of the fit values we do not observe a dependence of the string
tension on the strength of the external field. Therefore we do not show the
plot of the string tension {\it vs.} the external magnetic flux.

Summarizing, both magnetic and electric fields do not influence the string
tension. This fact is analyzed in terms of the monopole properties in
Section~\ref{sec:monopole:properties}. Before doing so we discuss the phase
structure of the model using the  Polyakov loop expectation value and the
corresponding susceptibility.

\section{Phase structure}
\label{sec:phase}

The confinement--deconfinement phase transition is usually detected using
the Polyakov loop vacuum expectation value. It is convenient to study
$v.e.v.$'s of the powers of the  ``bulk Polyakov loop'' defined as follows:
\beqn
  \la |L| \ra = \frac{1}{L^2_s} \,
  \la |\sum\limits_{\mathbf x} L(\mathbf x)| \ra\,, \quad
  \la |L|^2 \ra = 
\frac{1}{L^2_s} \,
  \la |\sum\limits_{\mathbf x} L(\mathbf x)|^2 \ra = 
\frac{1}{L^2_s} \,
  \la \sum\limits_{{\mathbf x}, {\mathbf y}} L({\mathbf x}) L^+({\mathbf y})
  \ra \,. \label{bulk:pol:loop}
\eeqn
The Polyakov loop susceptibility is expressed via these quantities:
\beqn
\chi_L = {\la |L| \ra}^2 - \la  |L|^2 \ra \,.
\label{bulk:pol:sus}
\eeqn
According to the free energy arguments in the deconfinement phase the 
quantity $|L|$ should be non--vanishing while in the confinement phase 
this quantity becomes small (it approaches zero in the infinite volume limit).

In the following we show that the expectation value of the Polyakov loop,
similar to the correlator discussed above, also gets a large classical
contribution due to the external electric field. Therefore it should not be
used in general as an order parameter to  probe the restoration of
confinement (except for some special values of the electric field).

\subsection{Electric field}

\begin{figure}[!htb]
  \begin{tabular}{cc}
  \epsfxsize=7.0cm \epsffile{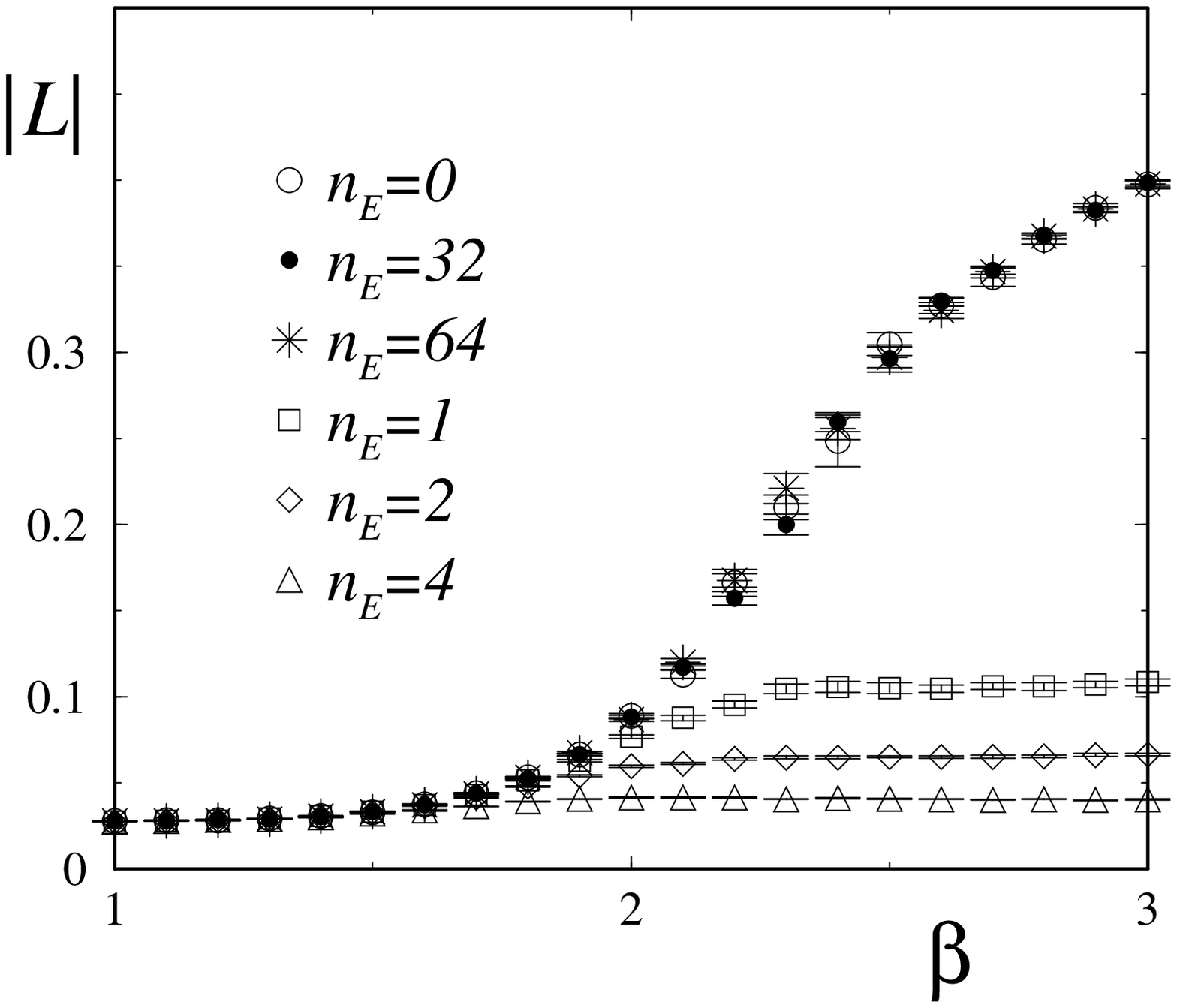}  &  
  \epsfxsize=7.0cm \epsffile{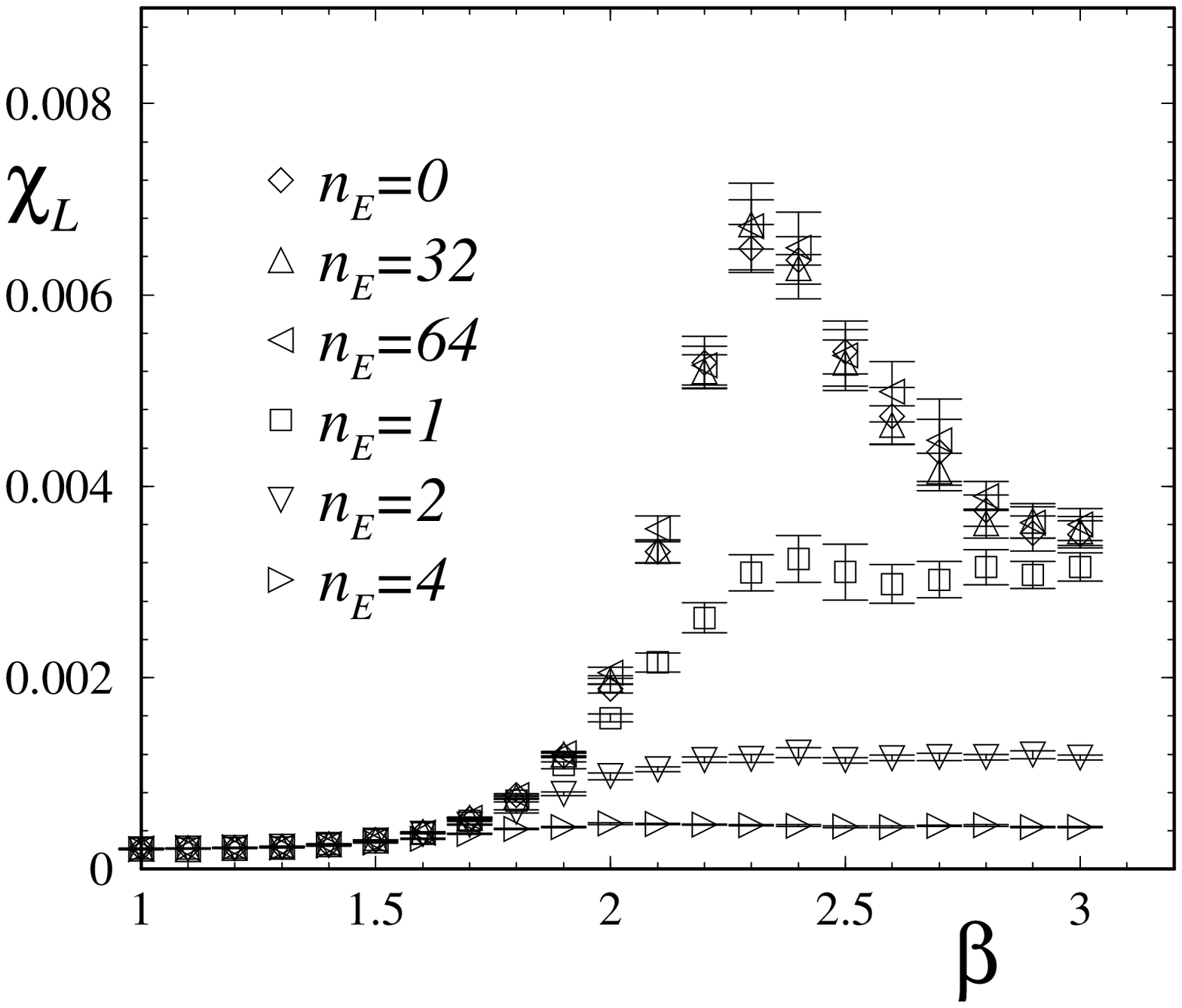} \\
  (a) & \hspace{1.5cm}  (b) \\
  \end{tabular}
  \caption{(a) The absolute value of the bulk Polyakov 
  loop~\eq{bulk:pol:loop} and (b) the Polyakov loop
  susceptibility~\eq{bulk:pol:sus} {\it vs.}
  $\beta$ for various values of the external electric field, $n_E$.}
  \label{fig:bulk:polyakov:loop:next}
\end{figure}

The behavior of the Polyakov loop {\it vs.} $\beta$ at various values of 
the external electric field is shown in
Figure~\ref{fig:bulk:polyakov:loop:next}(a). 
The plot of the Polyakov loop at zero field agrees with general expectations. 
However, as the electric field is turned on, the vacuum expectation value of
the Polyakov loop is decreasing. One might conclude that the external
electric field restores the  confinement phase (while being in deconfinement
at $n_E=0$) which is in clear contradiction with the results for the string
tension shown in the  previous Section. We remind the reader that the
external electric field  modifies the properties of the Polyakov loop
correlator classically. In particular, for not too large $n_E$ it enhances 
what we called the {\it effective} string tension measured from the Polyakov
plane--plane correlators perpendicular to the electric field ({\it cf.}
eq.~\eq{eff:sigma}). But, as we noted above, this effective string tension
does not describe  the confining properties of the system. 

Thus we may conclude that the rapid decrease of the Polyakov loop as function
of $n_E$  also does not mean that confinement is restored. 
Note that at large special values of the internal flux $\nint$,
eq.~\eq{cond:N:E}, the module of the Polyakov loop and the susceptibility 
(Figure~\ref{fig:bulk:polyakov:loop:next}(b)) as function of $\beta$ do not
differ from the zero field case.

Let us estimate the classical correction to the Polyakov loop expectation
value due to the external electric field $E$. For this purpose 
we consider the squared modulus of the Polyakov line, eq.~\eq{bulk:pol:loop}.
In the presence of $E$  it can be written as follows:
\beqn
  {\la |L|^2 \ra}_E = {\mathrm{const}}\cdot 
  \sum\limits_{x,y} {\mathrm e}^{2 \pi i \, n_{\mathrm{int}} \, x \slash L_s} 
  \, \la L(0,0) L^*(x,y) \ra\,. 
  \label{bulk:pol:loop:th:0}
\eeqn
Taking into account eq.~\eq{mitr} for the point--point Polyakov loop
correlator and summing over all momenta $p_i$ in this equation we
get\footnote{ According to our considerations above the dominating internal
flux is equal to the external flux in the vicinity of the phase transition and
thus we can safely put $\nint = n_E$.}:
\beqn
  {\la |L|^2 \ra}_E = \sum\limits_{m\ge 0} \frac{C_m}{
  \cos (2 \pi \nint \slash L_s) - \cosh(\sigma_m L_t)}\,.
  \label{bulk:pol:loop:th}
\eeqn
Here the expansion over the excited mass states $\sigma_m$ is written
explicitly and the lowest state corresponding to the string  
tension: $\sigma_0 = \sigma$. From
Figures~\ref{fig:bulk:polyakov:loop:next} it follows that
$\chi_L \ll \langle |L|\rangle^2$ for all $\beta$ and $n_E$ values.
Thus the square root of eq.~\eq{bulk:pol:loop:th} can serve as a good
estimator  for the Polyakov loop quantum average $\la | L | \ra$. 

\begin{figure*}[!htb]
  \begin{center}
    \begin{tabular}{cc}
    \epsfxsize=7.0cm \epsffile{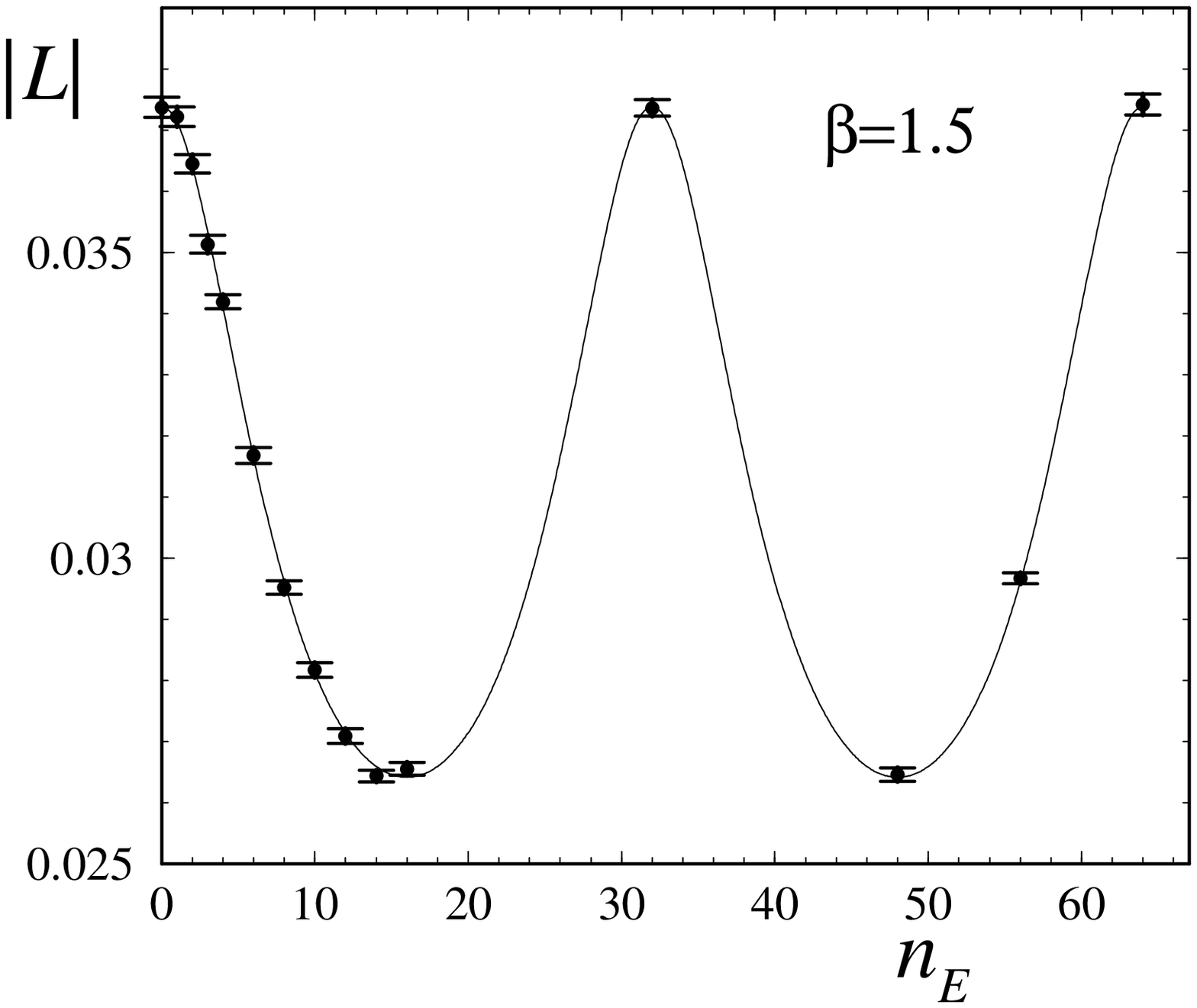}  &  
    \epsfxsize=7.0cm \epsffile{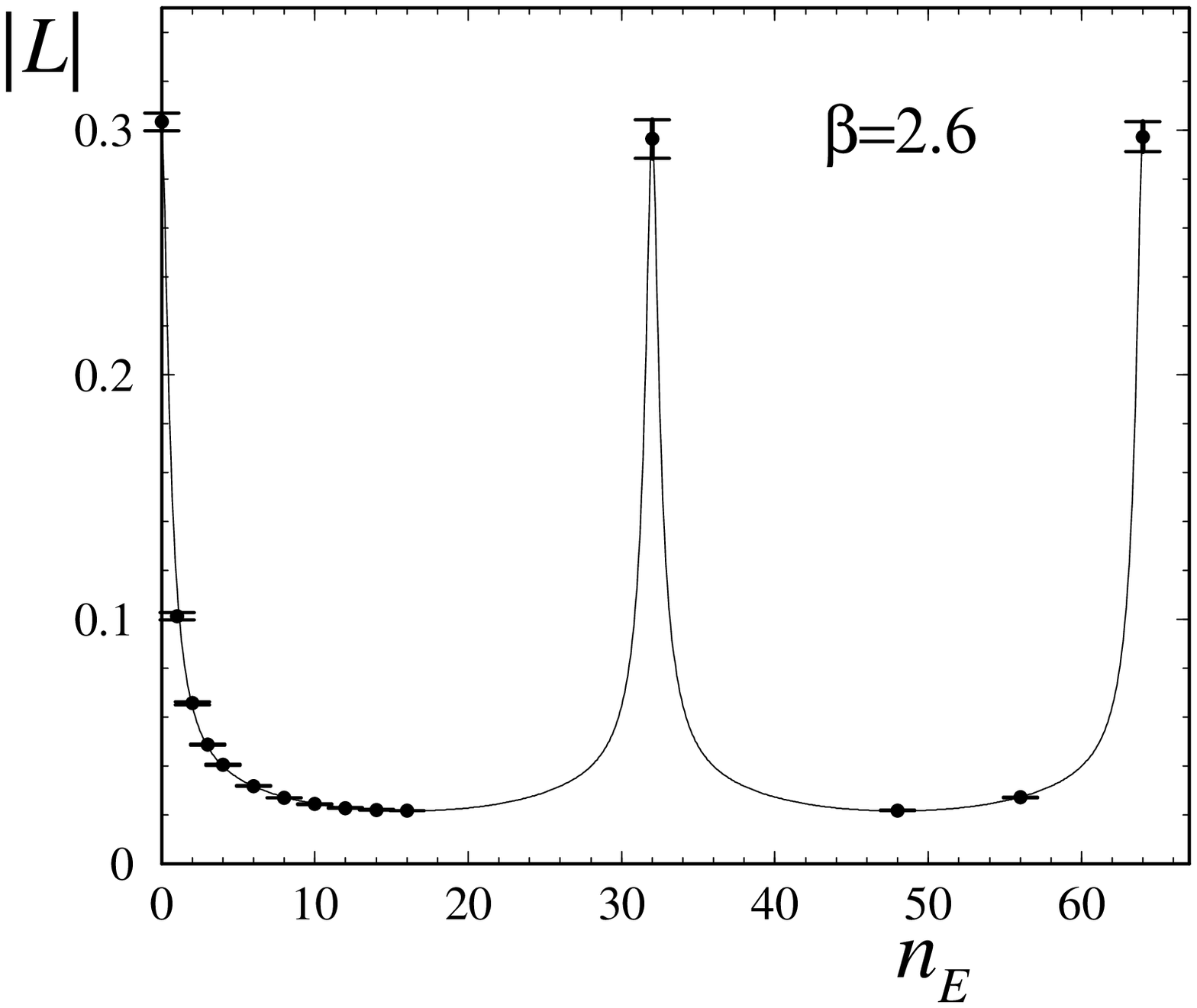} \\
    (a) & \hspace{1.5cm}  (b) \\
    \end{tabular}
  \end{center}
  \vspace{-0.5cm}
  \caption{The absolute value of the bulk Polyakov loop, \eq{bulk:pol:loop} 
  {\it vs.} the external electric flux $n_E$
  and its fit by formula~\eq{bulk:pol:loop:th} in (a) the confinement phase
  ($\beta=1.6$) and (b) the deconfinement ($\beta=2.5$) phase.}
  \label{fig:bulk:polyakov:fits:2}
\end{figure*}

We use only the first two terms of the expansion~\eq{bulk:pol:loop:th} to
fit the behavior of the {\it measured} Polyakov loop {\it vs.} the external
electric flux number. The corresponding fits for the confinement and
deconfinement phases are shown in
Figure~\ref{fig:bulk:polyakov:fits:2}(a,b), respectively. 
The best fits for the string tension in the confinement phase (at
$\beta=1.6$) give: $\sigma_0=0.167(13)$ and $\sigma_1=0.85(6)$. 
The value of $\sigma_0$ can be compared with the string tension without
external field,  $\sigma=0.170(4)$. These values agree with each other
within statistical errors. In the deconfinement phase ($\beta=2.5$) the same
picture is found: $\sigma_0=0.0096(5)$ and $\sigma_1=0.18(1)$, while the
independent measurements at zero field lead to $\sigma=0.0101(1)$. 
Therefore, the behavior of the Polyakov loop approximated via 
eq.~\eq{bulk:pol:loop:th} in the external electric field is consistent with 
the string tension measurements at zero--field.

So, in the case of non--vanishing external electric field, the bulk
Polyakov loop expectation value may vanish regardless of the value of the
actual string tension. Indeed, even  at $\sigma = 0$ (which is not
accessible on the finite lattice due to finite volume effects) the squared
Polyakov loop expectation value~\eq{bulk:pol:loop:th} is decreasing when
turning on the external electric flux. As we have mentioned in
Section~\ref{sec:string} the influence of the external field on the Polyakov
loop correlations is absent provided the condition~\eq{cond:N:E} 
for the internal flux $\nint$ is fulfilled. Now, a similar effect is
observed for the Polyakov loops themselves: the corresponding values for
$n_E = \nint= 32,\,64$ clearly coincide with $n_E = 0$ data. Thus we
conclude that the observables based on the Polyakov loops may have a 
usual physical sense only in the cases of the quantized internal
field~\eq{cond:N:E}.

\begin{figure*}[!htb]
  \begin{center}
    \begin{tabular}{cc}
      \epsfxsize=7.0cm \epsffile{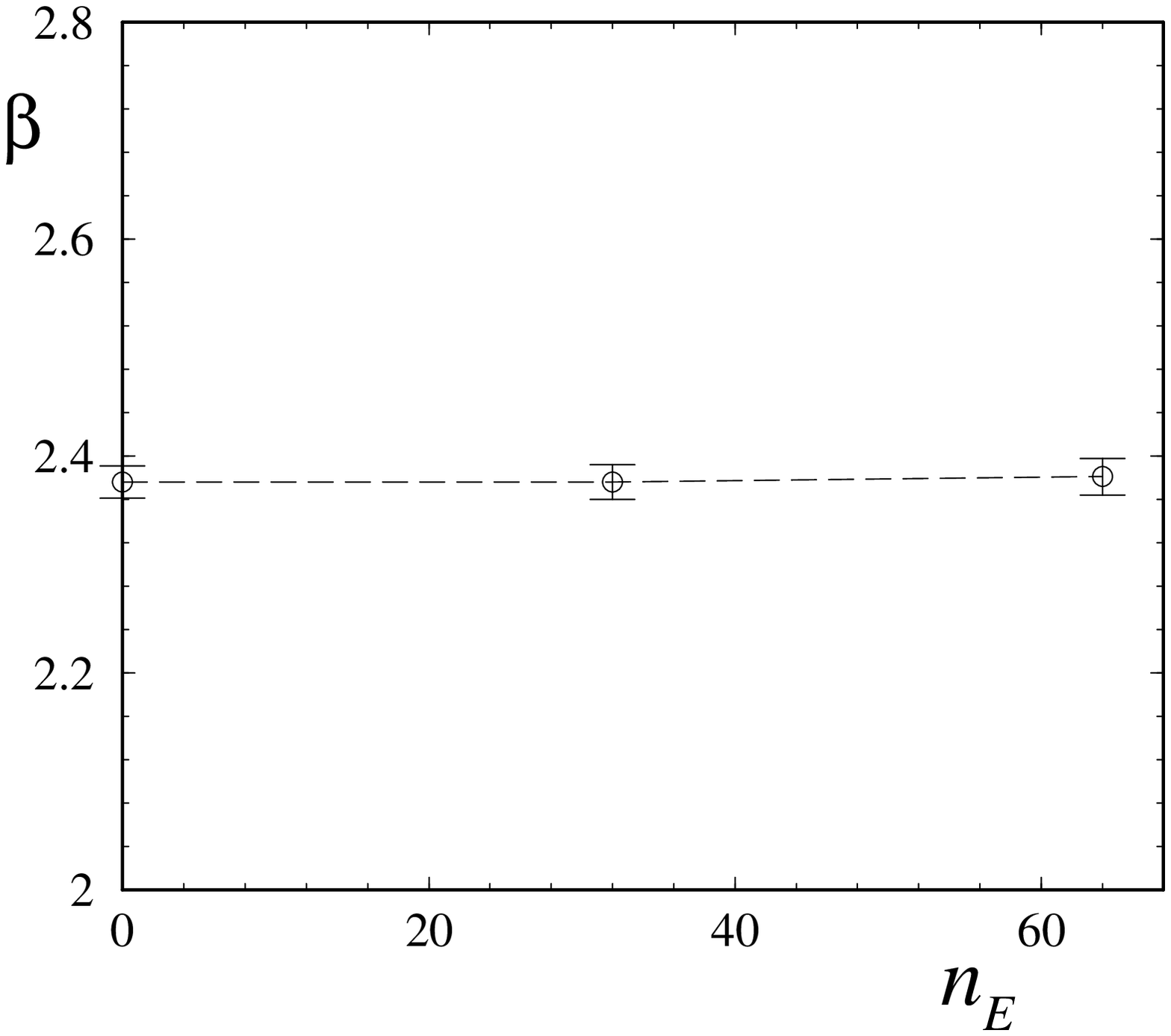}  &
      \epsfxsize=7.0cm \epsffile{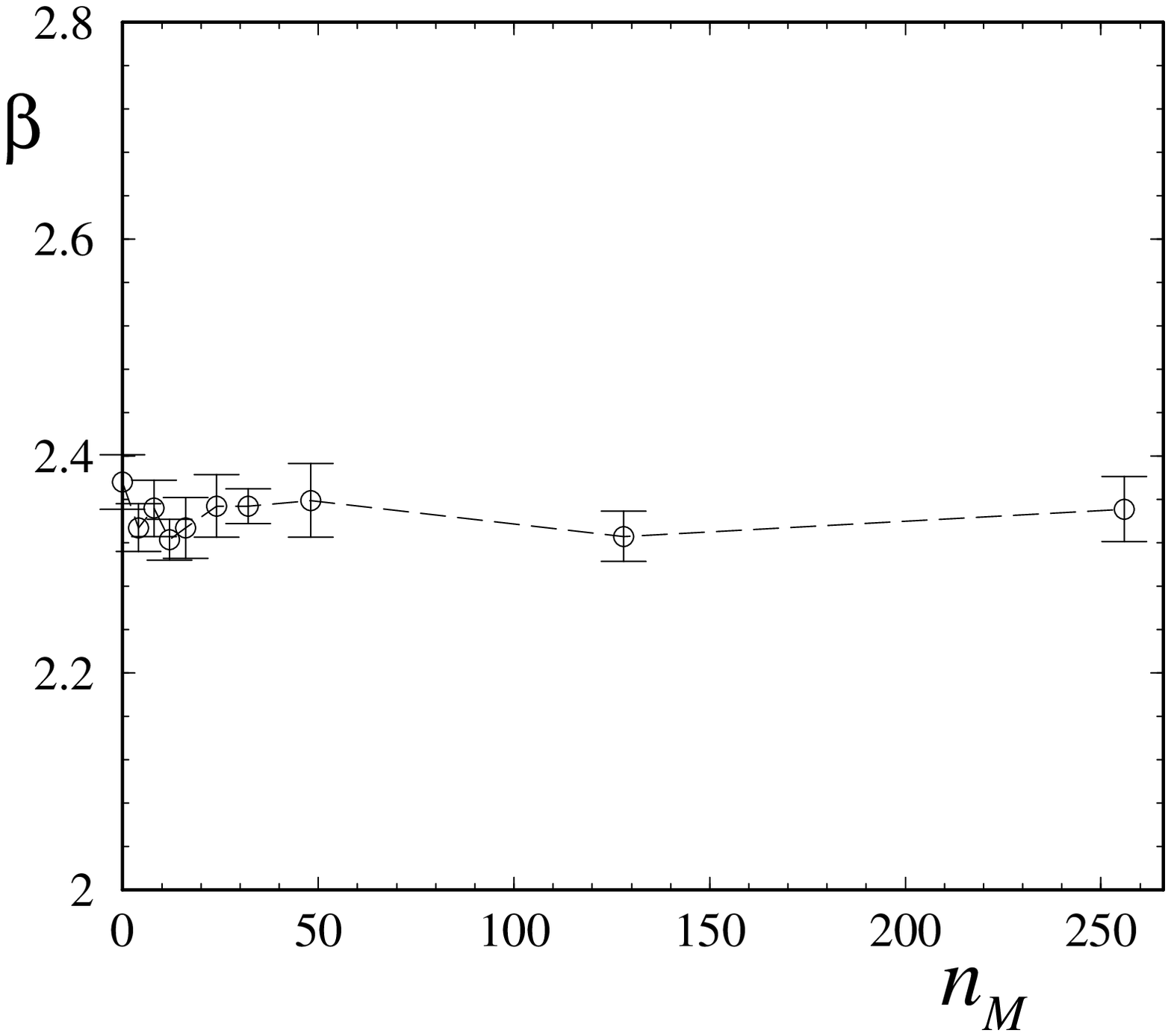} \\
      (a) & \hspace{1.5cm}  (b) \\
    \end{tabular}
  \end{center}
  \vspace{-0.5cm}
  \caption{Phase diagrams in the $n_{E/M} - \beta$ plane derived from the
   module of the Polyakov loop for (a) electric and (b) magnetic fields,
   respectively.}  
  \label{fig:phase}
\end{figure*}

Analogously, Figures~\ref{fig:bulk:polyakov:loop:next}(a,b) show that the
dependence of the Polyakov loop and its susceptibility on the coupling 
constant $\beta$ are the same (within errors) for $n_E = 0\,,32,\,64$.
The peaks in susceptibility may serve as good indicator of the
(pseudo)critical coupling constant. At these distinguished values of the
external electric field we have fitted the susceptibility near its maximum by
the following function:  
\beqn
  \chi^{\mathrm{fit}}_L(\beta) = \frac{c^2_1}{c^2_2 + (\beta - \beta_c)^2}\,,
\label{chi}
\eeqn
where $c_i$ and $\beta_c$ are fitting parameters. The results are shown in
Figure~\ref{fig:phase}(a) 
in the $ n_E - \beta$ plane. One can see that the influence of the external
field on the critical temperature is negligible.

\subsection{Magnetic field}

\begin{figure*}[!htb]
  \begin{center}
    \begin{tabular}{cc}
      \epsfxsize=7.0cm \epsffile{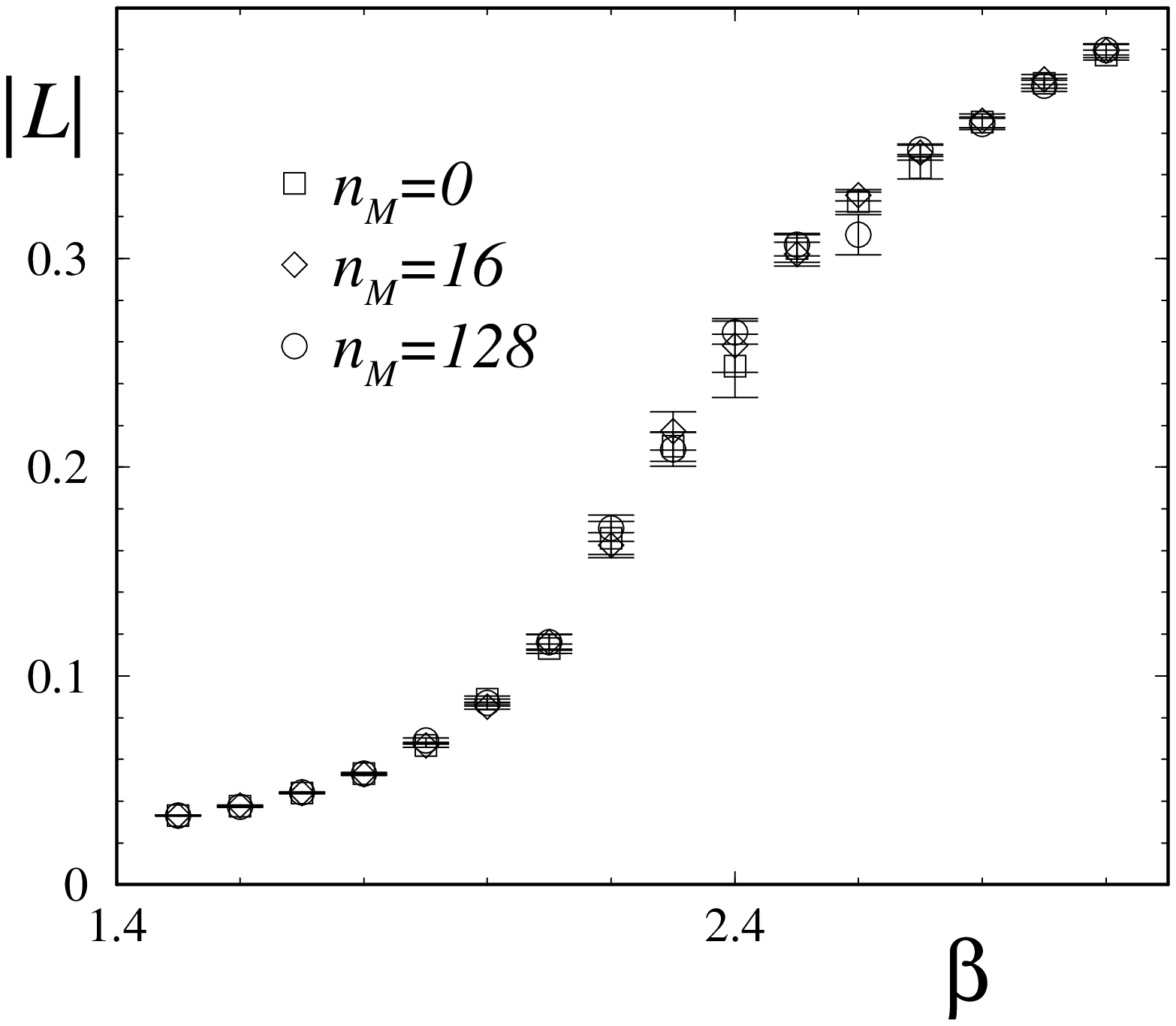} & 
      \epsfxsize=7.0cm \epsffile{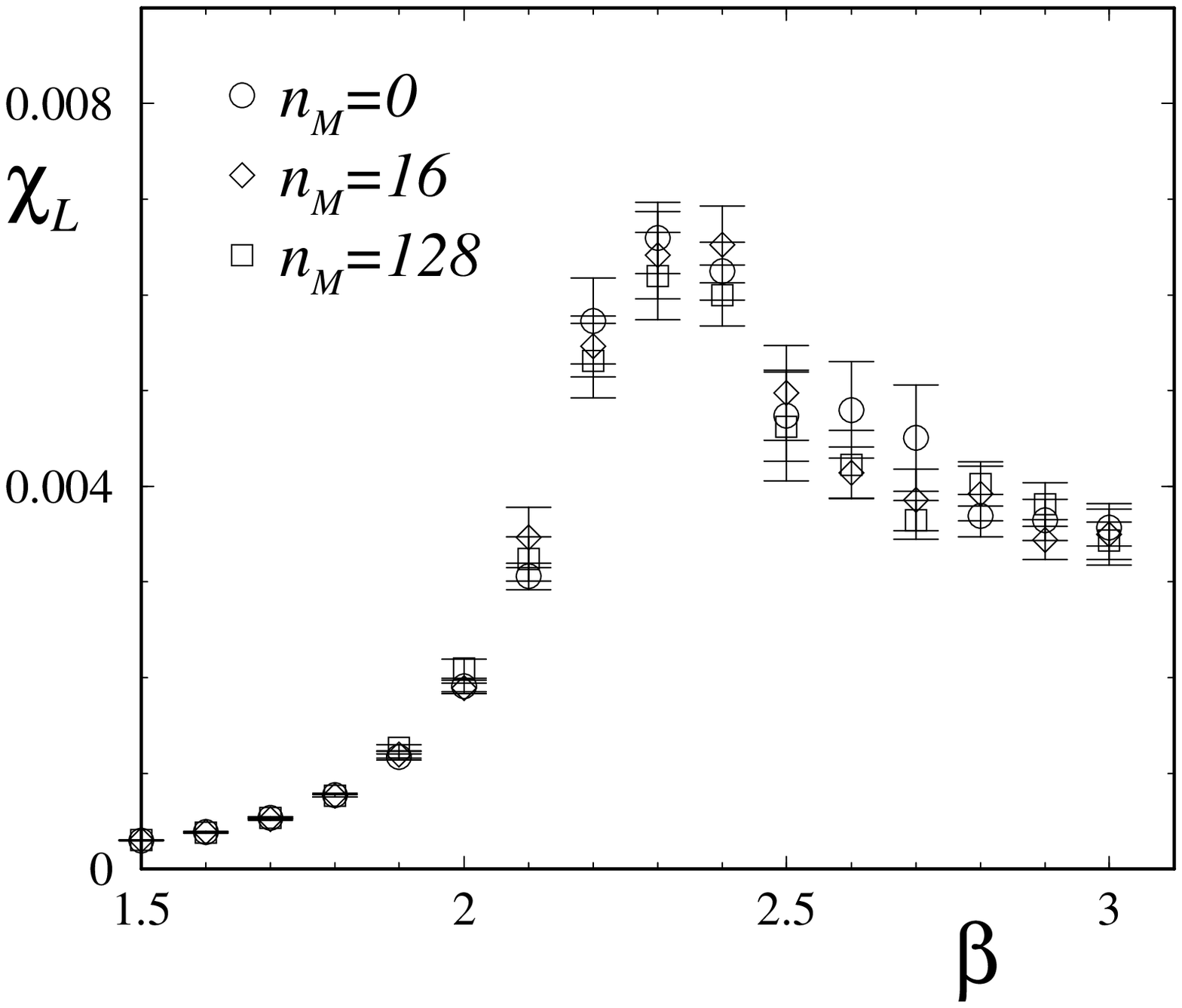} \\
        (a) & \hspace{1.5cm}  (b) \\
    \end{tabular}
  \end{center}
  \vspace{-0.5cm}
  \caption{(a) The Polyakov loop and (b) its susceptibility at various values
  of the external magnetic field $n_M$ {\it vs.} $\beta$.}
  \label{fig:polyakov:true:mag}
\end{figure*}

The tree level contribution to the Polyakov loop observables is absent in the
case of the external magnetic field. This is confirmed by
Figures~\ref{fig:polyakov:true:mag}(a,b)  
where the Polyakov loop and its susceptibility are shown as functions of
$\beta$, respectively. The data points for all considered values of the
external magnetic flux coincide  with each other within errors. We fit the
Polyakov loop susceptibility by eq.~\eq{chi} to get the (pseudo)critical
couplings. The corresponding phase diagram is shown in
Figure~\ref{fig:phase}(b) in the $n_M - \beta$ plane. One sees that the phase
transition points are insensitive to the strength of the external magnetic
field.

It is interesting to note that in non--Abelian gauge theories in $3+1$
dimensions the situation is quite different~\cite{Cea}. Here the  
position of the deconfining phase transition depends strongly on the strength
of the applied magnetic field. The simplest, somewhat naive explanation of
this fact could be as follows. In the non--Abelian gauge theory the coupling
between gauge fields is stronger than in the Abelian theory. In particular,  
there is a correlation between spatial and temporal components of the fields 
which may imply that the external magnetic field (encoded in the spatial
components) induces internal electric fluxes on the quantum level. As we
have seen the electric fluxes strongly influence the Polyakov loop correlators
generating the effective string tension~\eq{eff:sigma}. Contrary to the
Abelian case this influence is not a merely classical (or, just inherent to
the way of introducing the external field) but a real quantum effect.

In the next Section we study the effects of the external fields on the
agents of confinement, the Abelian monopoles, to explain the observed
behavior of the system.  A similar study for the non--Abelian gauge theory
is underway~\cite{CIS2001c}.

\section{Monopole properties}
\label{sec:monopole:properties}

The basic quantity to describe the behavior of the monopoles is the
monopole density, 
\beqn
  \rho_{\mathrm{mon}} = \sum_c |m_c|\,, 
\eeqn
where $m_c$ is the integer valued monopole charge inside the cube $c$ defined
in the standard way~\cite{DGT}:
\beqn
  m_c = \frac{1}{2\pi} \sum\limits_{P \in \partial c} {(-1)}^P \, {[\dd
  \theta]}_{\mathrm{mod} \, 2 \pi}\,.
\eeqn
In our previous study~\cite{CIS2001a} we have demonstrated that the monopoles
are sensitive to the phase transition in the compact Abelian gauge model at
finite temperature. Although in $(2+1)D$ we have magnetic and electric field
among the three components of the field strength tensor, the sources of the
respective fluxes will be simply called ``monopoles'' or ``magnetic charges''
in the following. 

The mechanism which drives the finite temperature deconfinement phase
transition is monopole binding. In the zero temperature case the plasma of
monopoles and anti--monopoles can explain the permanent confinement of
oppositely charged electric test charges~\cite{Polyakov} in bound states, kept
together by a linear potential. Confinement appears due to the screening of
the magnetic field induced by the electric current circulating along the
Wilson loop. Monopoles and anti--monopoles form a polarized sheet of finite
thickness (``string'') along the minimal surface $|A|$ spanned by the Wilson
loop. The formation of the string leads, for non--vanishing electric current,
to an excess of the free energy  equal to $\sigma |A|$.

At finite temperature, dipoles are formed both in the confinement and
deconfinement phases. In the deconfinement phase tightly bound dipoles
dominate in the vacuum.  The dipole plasma is inefficient to completely screen
the field created by the electric currents running along the pair of Polyakov
loops. This explains the absence of confinement in this phase.

Besides measuring the density of monopoles, we have studied the properties of
the monopole ensembles  by investigating the structure of monopole clusters.  
Clusters are connected groups of monopoles and anti--monopoles, where each
object is separated from at least one neighbor belonging to the same
cluster by a distance less or equal than $R_{\mathrm{max}}$. In the
following we use $R_{\mathrm{max}}=\sqrt{3}~a$ which means that neighboring
monopole cubes should share at least one single corner\footnote{In
Ref.\cite{TeperSchram} a similar definition  has been used to investigate
tightly packed clusters with $R_{max}=a$. In our case the condition for the
cluster is more relaxed.}. The increase of the coupling constant leads not
only to an increase of the temperature, eq.~\eq{temp}, but also to a
decreasing lattice spacing $a$, eq.~\eq{beta}. Thus at different $\beta$ the
same characteristic distance  $R_{\mathrm{max}}$ corresponds to different
physical scales. Therefore our results presented here are of a qualitative
nature.

In our study without external field we have found that the dipoles
are oriented dominantly in the temporal direction. At the confinement phase
transition mostly clusters with two constituents or single monopoles  and
anti--monopoles were observed. Decreasing further the temperature (or
$\beta$), the monopoles become dense and start to form connected clusters (on
a coarser and coarser lattice) containing various numbers of monopoles and
anti--monopoles. The largest clusters have been found to be more and more
spherical.

Finally we observed that only charged monopoles clusters in the plasma 
(mainly individual monopoles) are needed to explain the measured string
tension and, therefore, are responsible for confinement. 

\subsection{Electric field}

\begin{figure*}[!htb]
  \begin{center}
    \begin{tabular}{cc}
      \epsfxsize=7.0cm \epsffile{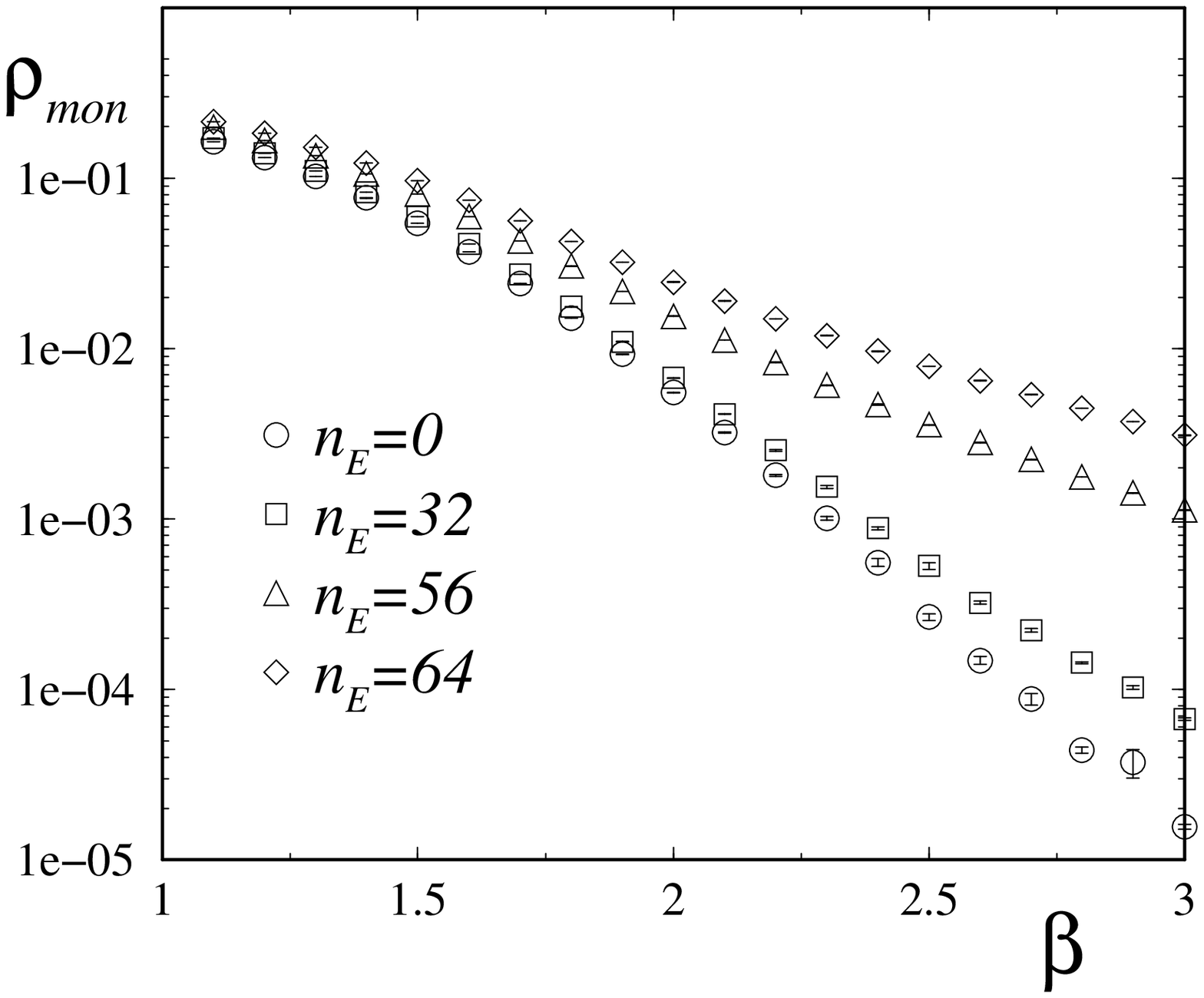} & 
      \epsfxsize=7.0cm \epsffile{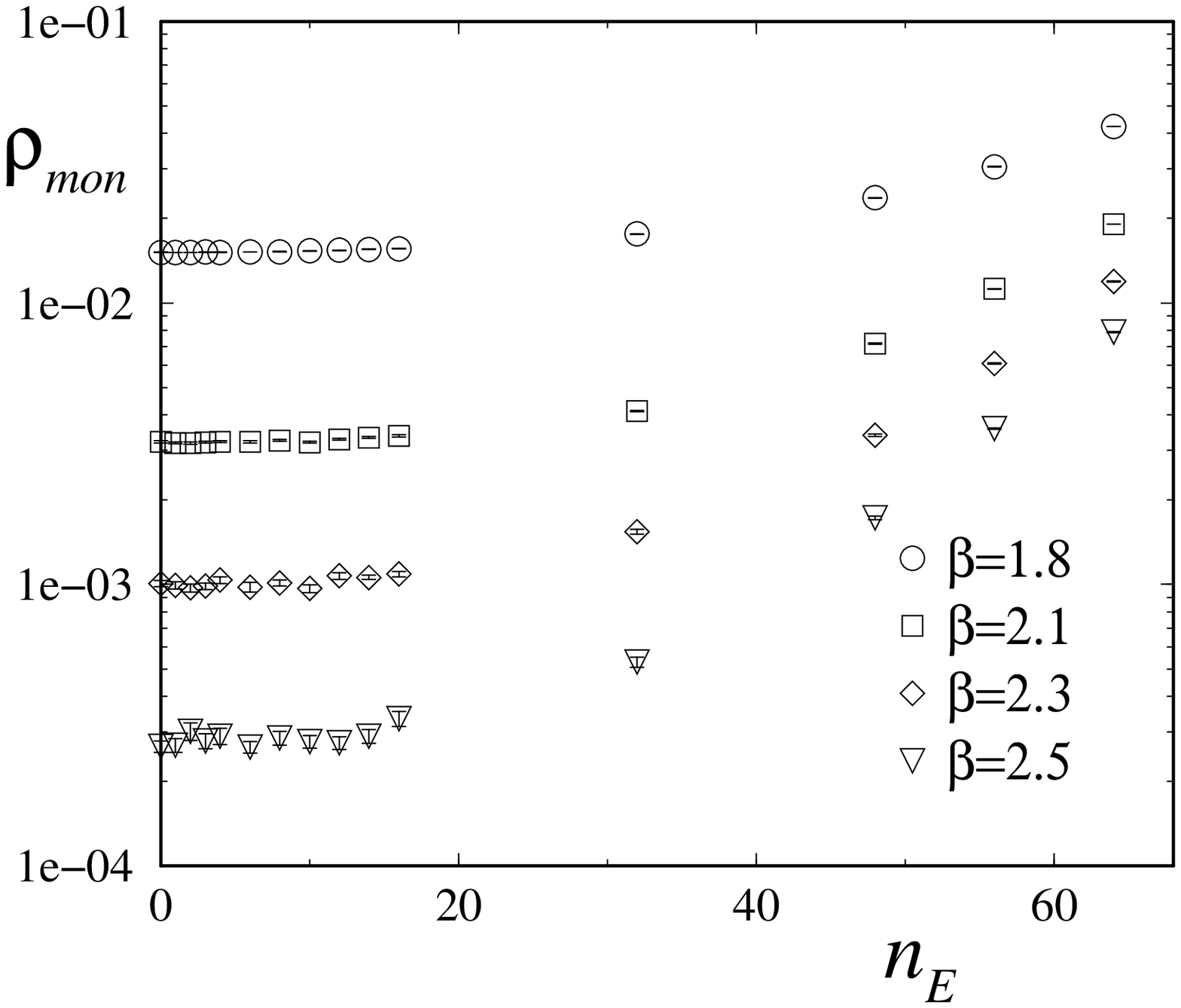} \\
      (a) & \hspace{1.5cm}  (b) \\
    \end{tabular}
  \end{center}
  \vspace{-0.5cm}
  \caption{Monopole density $\rho_{\mathrm{mon}}$ (a) {\it vs.} $\beta$ for
  various external electric fluxes, $n_E$; 
  (b) {\it vs.} $n_E$ for various $\beta$ values.} 
  \label{fig:mon:next}
\end{figure*}

To begin, we plot in Figure~\ref{fig:mon:next}(a) 
the total monopole density $\rho$ as function of the coupling $\beta$ for
various values of the external electric flux $n_E$. The monopole density is a
decreasing function of $\beta$ at any value of the external field. However,
$\rho$ increases as the function of the strength of the applied external
field, Figure~\ref{fig:mon:next}(b). The effect of the external field is very
essential: the monopole density is increased up to almost two orders of
magnitude (depending on the temperature) for the largest external flux
values (compared to the system at zero external field). 

\begin{figure*}[!htb]
  \begin{center}
    \begin{tabular}{cc}
      \epsfxsize=7.0cm \epsffile{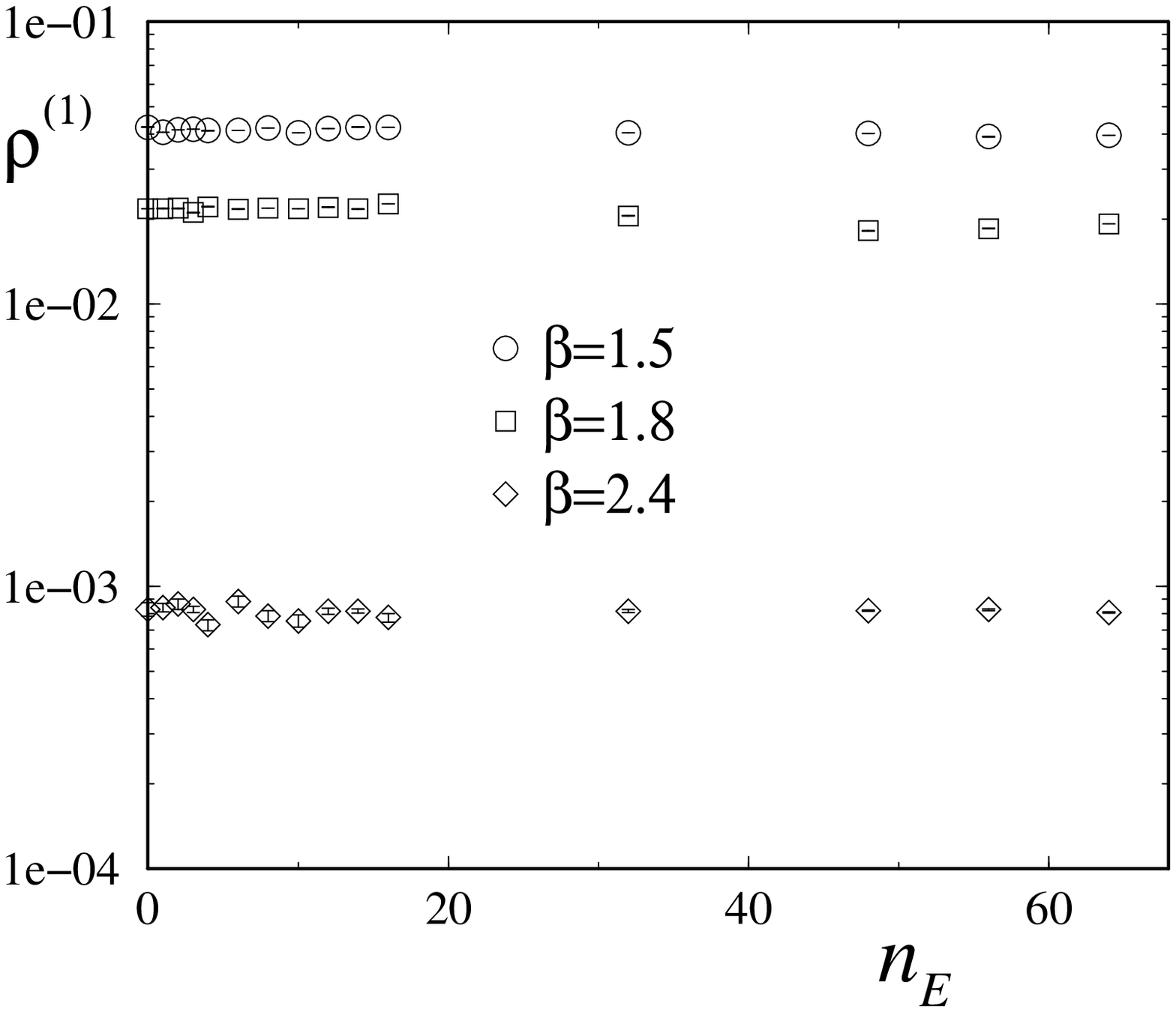} & 
      \epsfxsize=7.0cm \epsffile{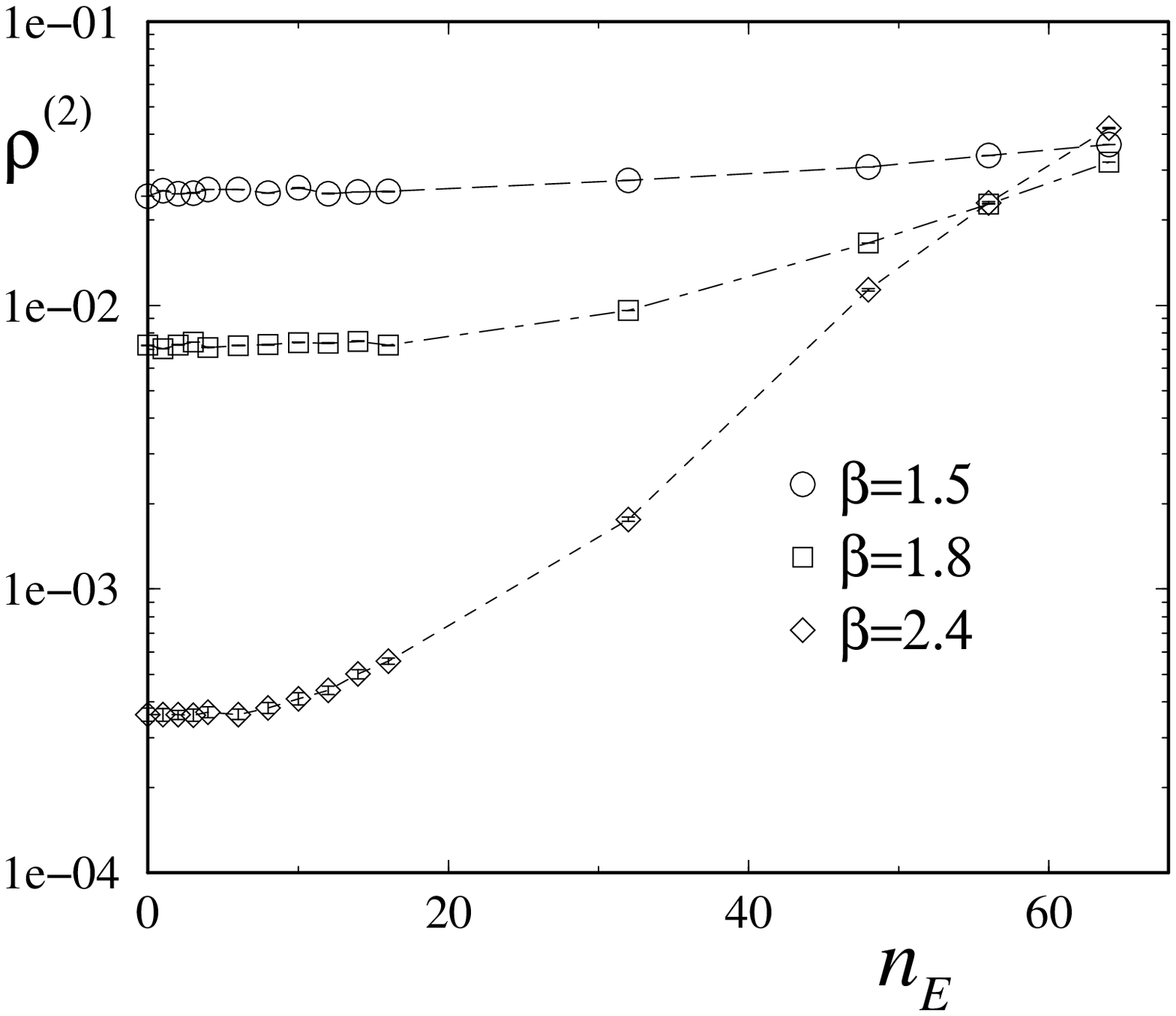} \\
       (a) & \hspace{1.5cm}  (b) \\
    \end{tabular}
  \end{center}
  \vspace{-0.5cm}
  \caption{The density of clusters of size (a) $N=1$ (single
 monopoles and anti--monopoles) and (b) $N=2$ (dipoles) 
 {\it vs.} external electric flux $n_E$ at various values of $\beta$.} 
  \label{fig:clust}
\end{figure*}

However, an increased total monopole density does not mean in general that
the confining properties of the system are enhanced.  Only charged monopole
clusters in the plasma state contribute to the string tension between
electrically charged test particles. Tightly bound monopole pairs are not
expected to contribute to the string tension. Therefore, we use our cluster
labeling algorithm to look into the structure of the monopole ensemble. 
In Figures~\ref{fig:clust}(a,b) 
we show how the densities of single monopoles (clusters made of just one
(anti--)monopole) and of dipoles (clusters made of two oppositely charged
objects), respectively, depend on the strength of the external electric
field. 

One clearly sees that the plasma component of the single monopole
ensemble does not feel the electric field at all. Thus the confining
properties of the system should not depend on the external electric field in
agreement with the conclusions made in the previous Sections.

On the other hand, the dipole density changes drastically with increasing
external field: the field creates the magnetic dipoles from the vacuum.  Note
that the larger the temperature (or, equivalently, $\beta$), the larger is
the increase of the dipole density. This fact is connected with the screening
of the external fields inside the medium discussed in
Section~\ref{sec:Polloop}. The larger the temperature the larger the fields
are inside the medium. As a result, the effect of the external field becomes
stronger with increasing temperature.

In a non--zero electric field the system is anisotropic in all directions.
The electric field is directed along the $y$ axis, the ``temperature''
direction, $z$, is influenced by compactification. Therefore, we have to
expect that the average sizes of the dipoles in different directions are not
the same. At zero or small external field the dipoles are mainly directed
along the temporal axis~\cite{CIS2001a}.

Increasing the external field, the dipoles are expected to become elongated
along the direction of the applied field. Moreover, we have observed in
Section~\ref{sec:Polloop} that the strength of the internal field inside the
medium relative to the external field increases as function of the coupling
$\beta$. Thus the elongation of the dipoles in the field direction should
increase with $\beta$.

\begin{figure*}[!htb] \begin{center}
  \begin{tabular}{cc} 
     \epsfxsize=7.0cm \epsffile{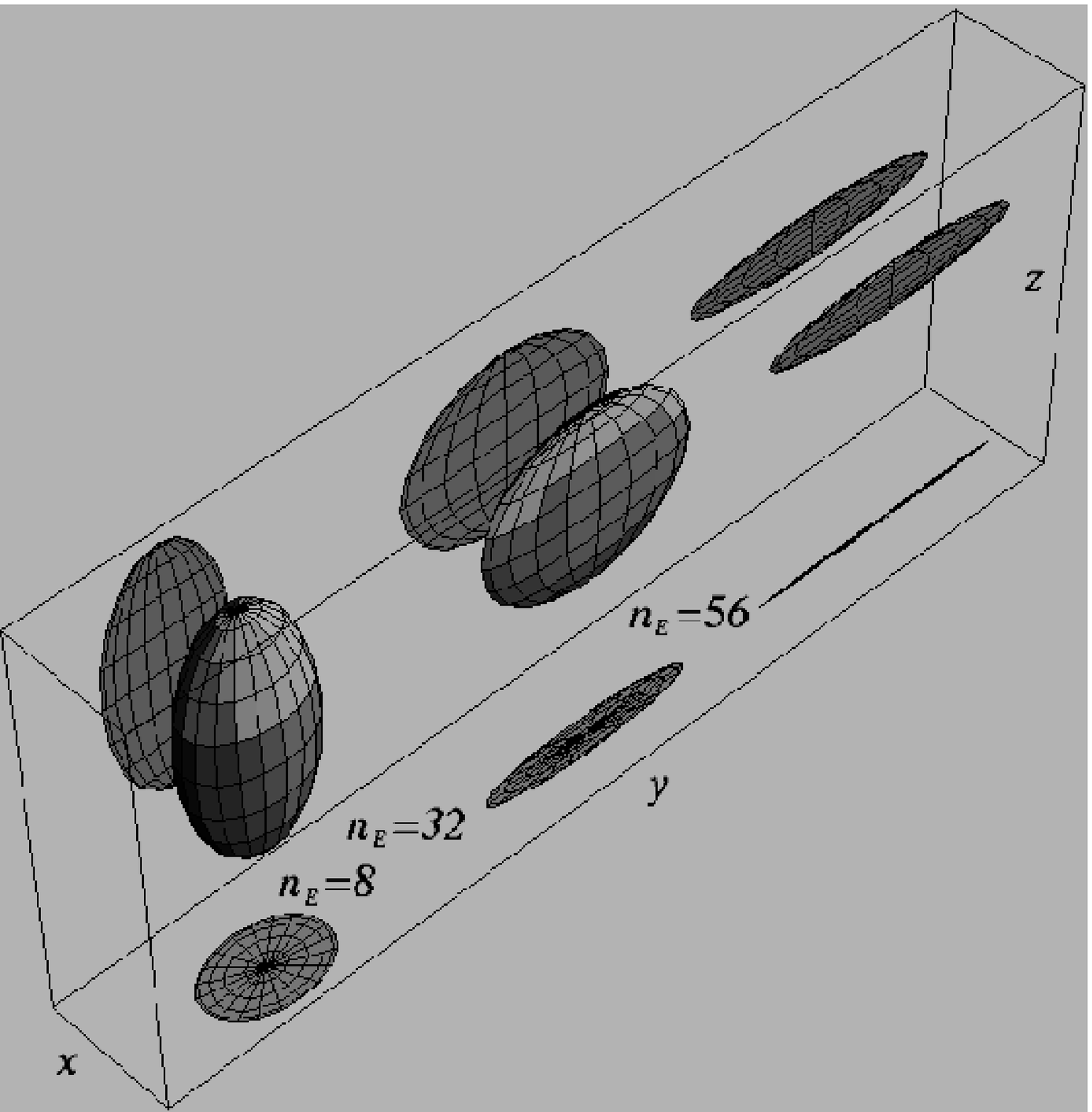} & 
     \epsfxsize=7.0cm \epsffile{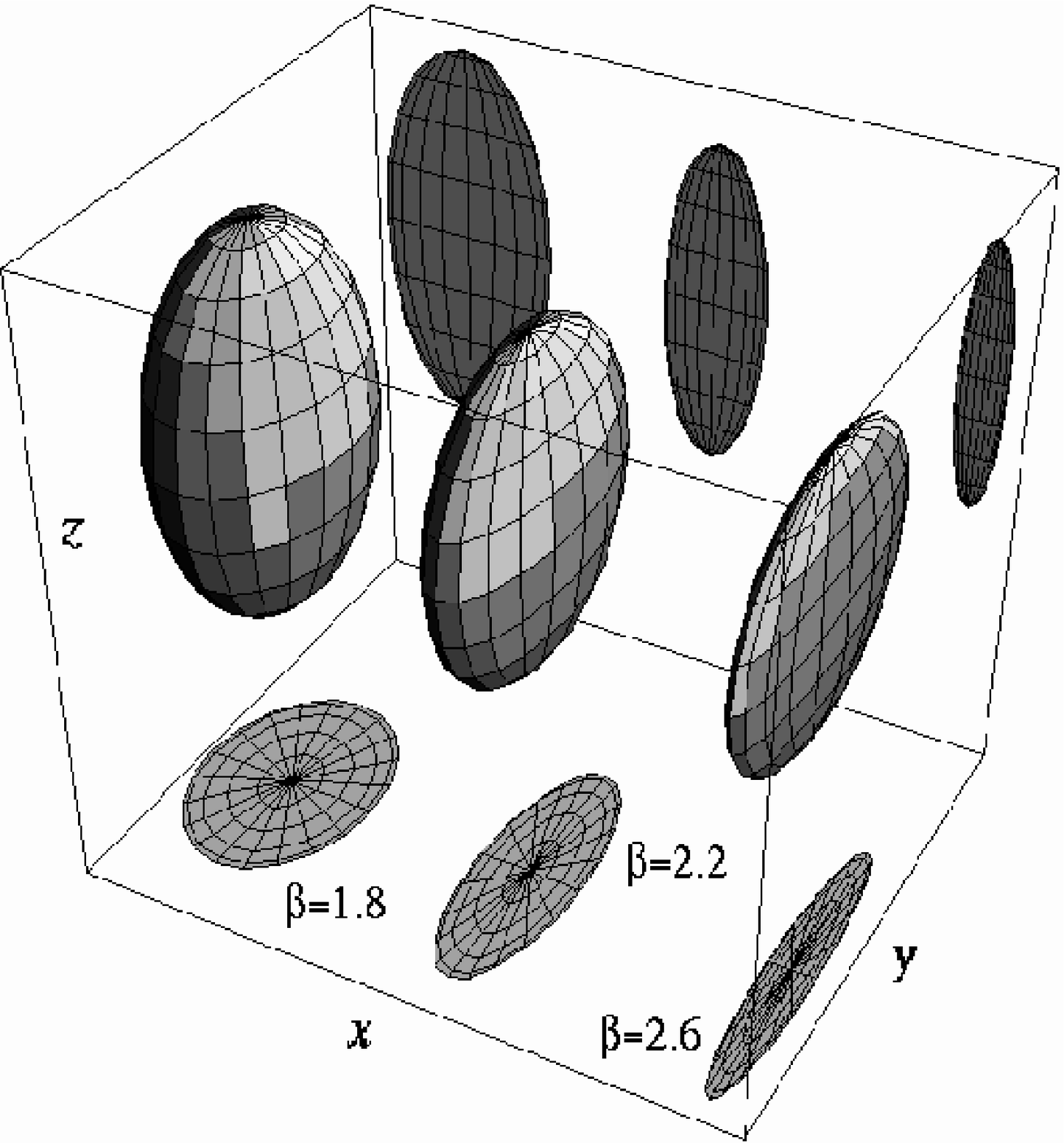} \\
      (a) & \hspace{1.5cm}  (b) \\
  \end{tabular}
  \end{center}
  \vspace{-0.5cm}
  \caption{The mean dipole anisotropy for increasing values of (a) the
  external electric flux $n_E$ at fixed $\beta=1.8$; (b)  $\beta$ at
  fixed $n_E$.} 
\label{fig:clust:3d}
\end{figure*}

All these effects are demonstrated in Figure~\ref{fig:clust:3d} 
using ellipsoids, the axes of which are equal to the average dipole sizes in
the $x,y$ and $z$ directions. In Figure~\ref{fig:clust:3d}(a) we show how the
mean dipole anisotropy changes with increasing external electric flux at fixed
$\beta=1.8$ (in the confinement phase). Figure~\ref{fig:clust:3d}(b)
demonstrates the dependence of the ellipsoids on $\beta$ at fixed flux $n_E$.
For convenience the projections of the ellipsoids onto the $x-y$ and $x-z$
planes are also shown. 

\subsection{Magnetic field}
\label{sec:monopole:properties:magnetic}

\begin{figure*}[!htb]
  \begin{center}
    \begin{tabular}{cc}
       \epsfxsize=7.0cm \epsffile{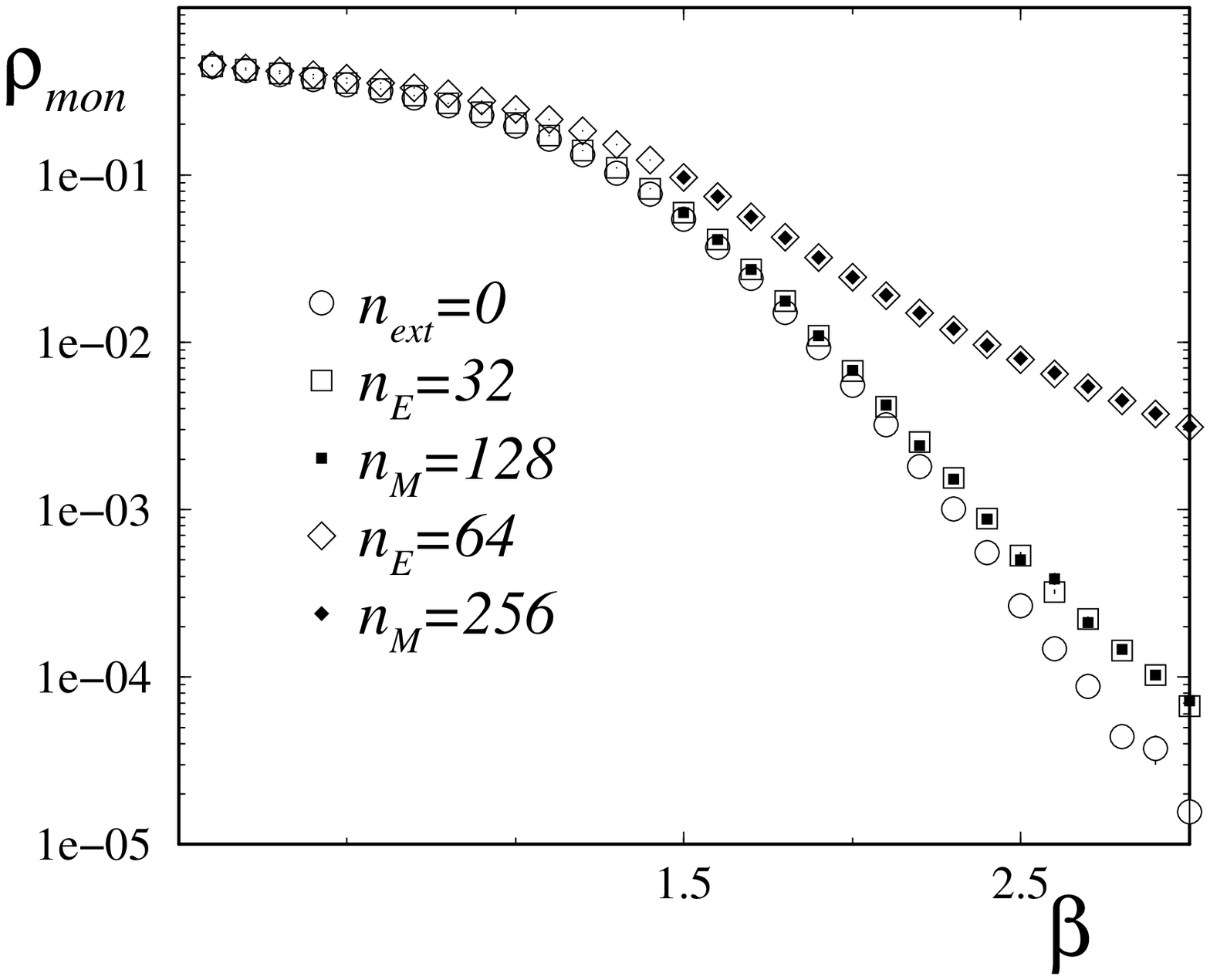} &
       \epsfxsize=6.5cm \epsffile{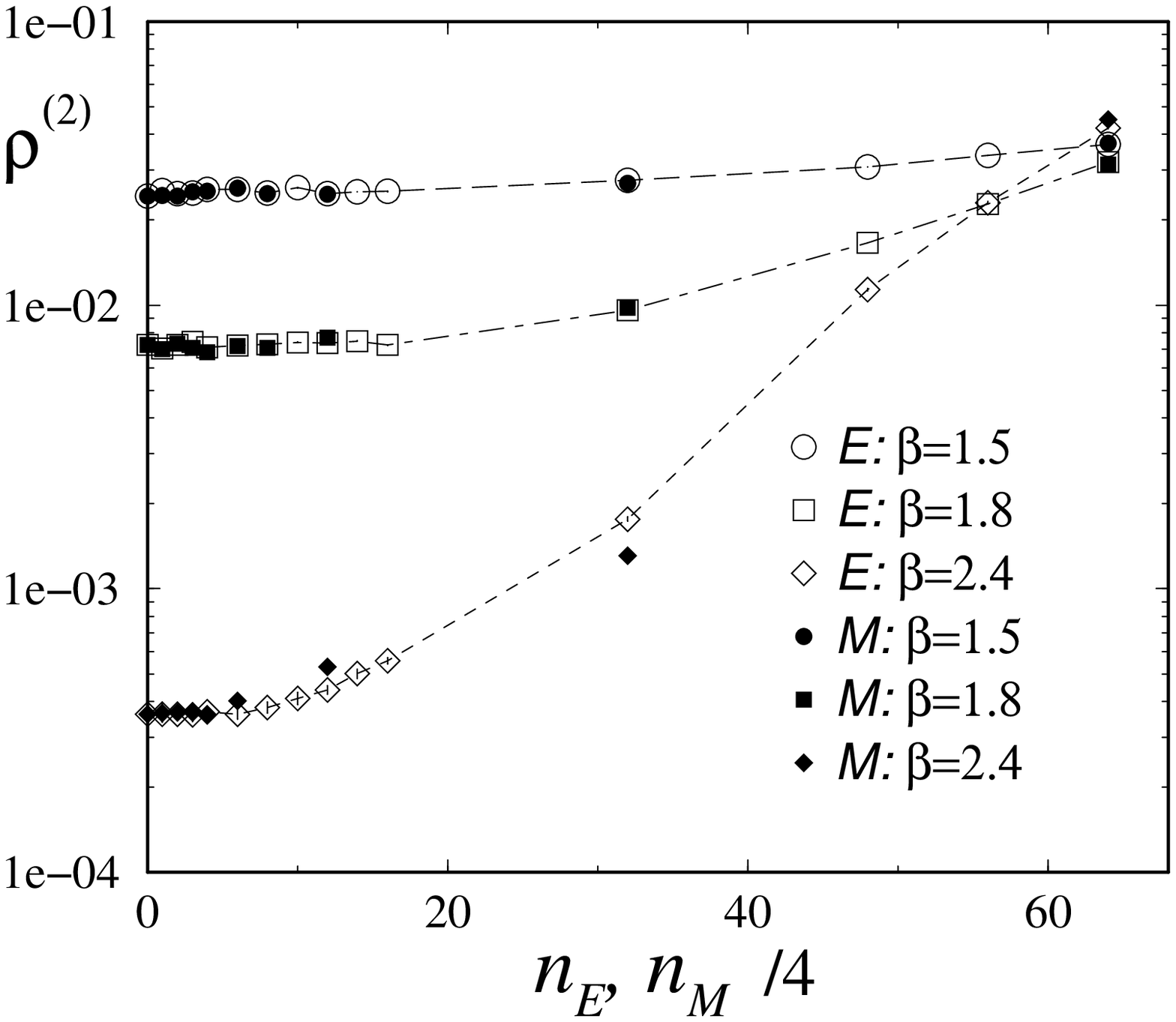} \\
       (a) & \hspace{1.5cm}  (b) \\
    \end{tabular}
  \end{center}
  \vspace{-0.5cm}
  \caption{(a) Total monopole density 
   $\rho_{\mathrm{mon}}$ {\it vs.} $\beta$ for
   various external flux values  $\next$.
  (b) density of dipole clusters {\it vs.} 
  $\next=n_{E} = n_M \slash 4$ for some $\beta$ values.}
  \label{fig:mon:3}
\end{figure*}

The influence of the magnetic field on the monopole densities is very
similar to that of the electric field. In Figures~\ref{fig:mon:3}(a,b)
we present the measured total monopole density and the dipole density, as
functions of $\beta$ and  the external fluxes, respectively. 
To compare the measurements for magnetic fluxes with those at non--zero
external electric field, we note that for our lattice geometry the strengths
of the magnetic and electric fields are equal to each other provided $n_M = 4
n_E$. According to those Figures both total monopole density and  dipole
density do not depend on the type of the external fields if the field
strengths are the same. We have checked that the single monopole density
coincides for magnetic and electric fields of the same strength as well. Thus
the cluster structure does not depend on whether the external field is of
electric or magnetic kind.

\begin{figure*}[!htb]
  \begin{center}
    \begin{tabular}{cc}
       \epsfxsize=6.5cm \epsffile{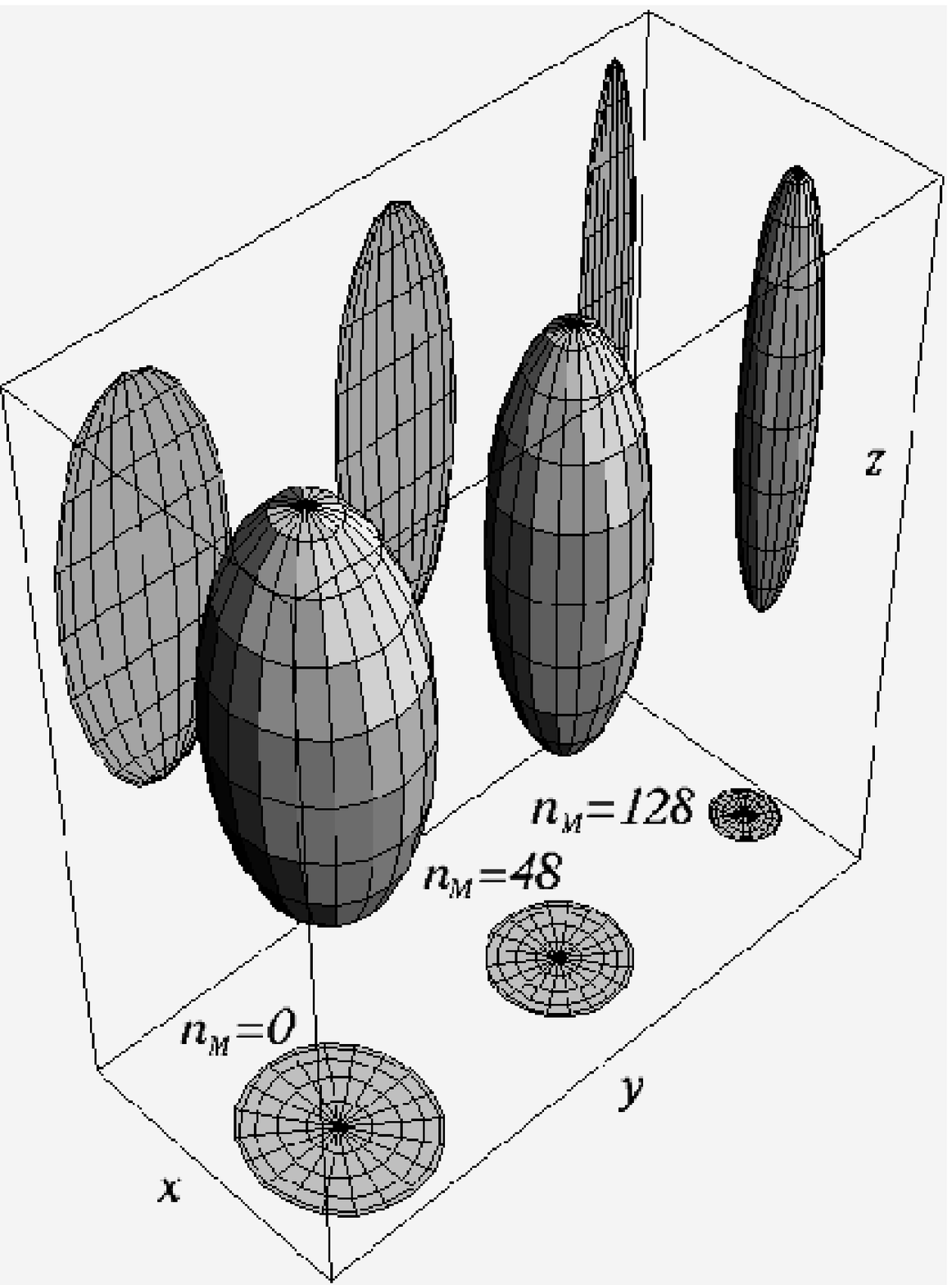} & 
       \epsfxsize=7.5cm \epsffile{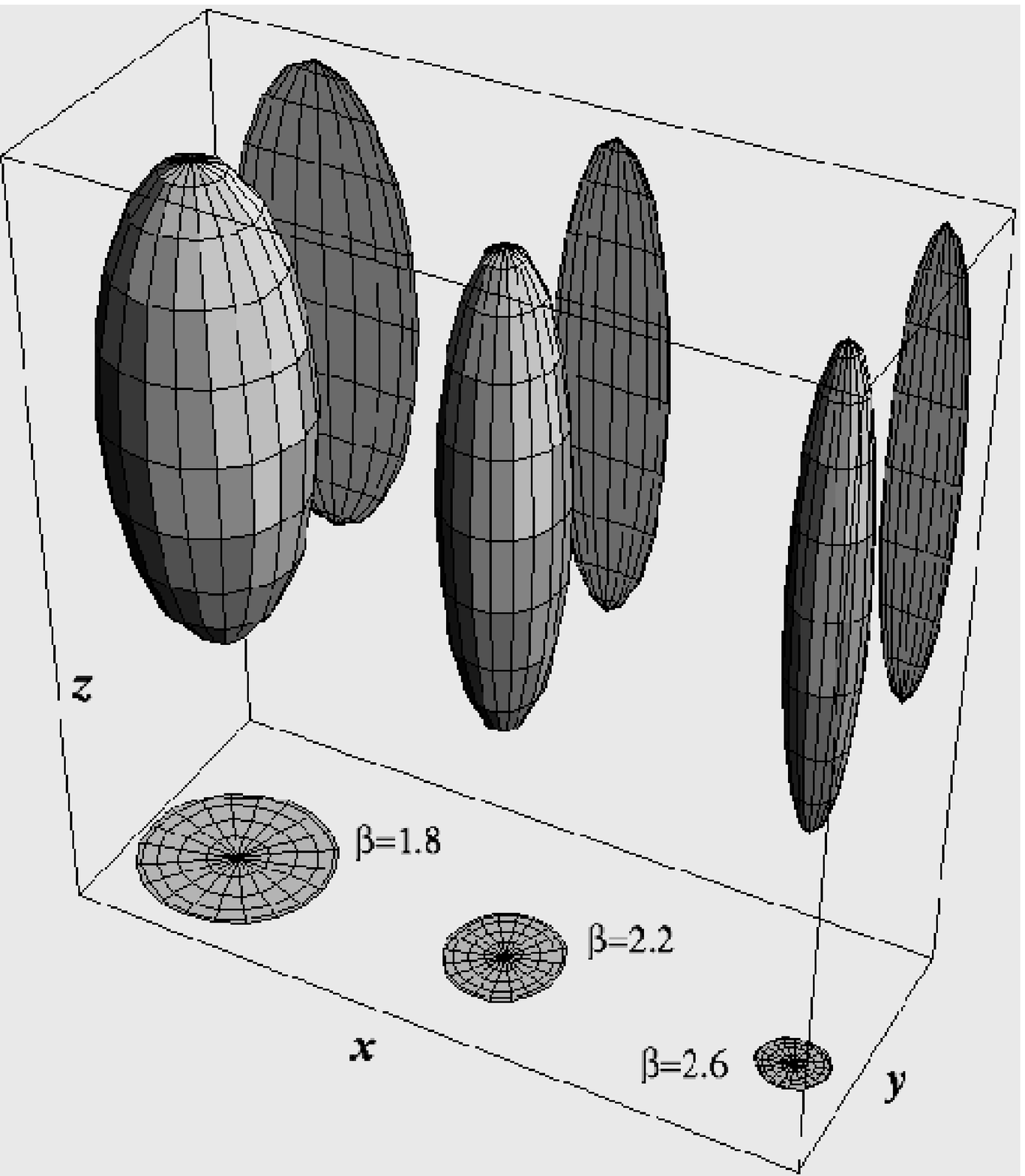} \\
       (a) & \hspace{1.5cm}  (b) \\
    \end{tabular}
  \end{center}
  \vspace{-0.5cm}
  \caption{The mean dipole anisotropy for increasing values of (a) the
    external magnetic flux $n_M$ at fixed $\beta=1.8$; (b)  $\beta$ at
   fixed $n_M$.}
  \label{fig:clust:m:3d}
\end{figure*}

The mean dipole anisotropy depends on the {\it type} (direction) of
the external field. We show the behavior of this quantity in
Figures~\ref{fig:clust:m:3d}(a,b)  
for increasing  external magnetic field and $\beta$, respectively. 
In the external magnetic field the system is symmetric in the $x$ and $y$
directions. The dipoles become more elongated in the (temporal) $z$ direction
for stronger external magnetic fields. Increasing the coupling $\beta$ leads
to a larger anisotropy since the medium does not screen the external field in
the deconfinement (large$\beta$).

The orientation properties of the dipoles influence the resulting
electromagnetic fields in the medium, Figure~\ref{fig:theta}.
According to Figures \ref{fig:clust:3d}(a) and \ref{fig:clust:m:3d}(a) 
the stronger the external field the larger elongation of the dipoles in the
direction of the field is. In turn this increases the dipole momenta in the
direction of the applied field and, as a result, leads to an enhanced
screening of the external fields by the media. This effect is clearly observed
in Figure~\ref{fig:theta}. 

Another interesting effect due to the dipole orientation is the clear
difference in screening for strong magnetic and electric fields of equal
strengths in the deconfinement phase, Figure~\ref{fig:theta}. According to
Figure~\ref{fig:mon:3}(b), the dipole densities for both  cases are the same.
Since $L_t < L_s$, the dipole magnetic moment density projected to the
magnetic field direction is larger than the corresponding quantity for the
electric field direction. On the other hand, the larger the dipole moment
density inside the medium the stronger the field attenuation is. Thus at
strong external fields the screening must be more effective for magnetic
fields compared to electric fields. This mechanism works at high
temperatures (large $\beta$) where the dipole fraction is dominant ($cf.$
Figures~\ref{fig:clust}(a,b)).

\section{Conclusion}

We have investigated the properties of the $3D$ compact electrodynamics at
finite temperature in external constant electric and magnetic fields. The
main result is that the deconfinement temperature is insensitive to both
electric and magnetic external fields. We have found the reason for this
behavior in terms of the monopole degrees of freedom: the external fields
create tightly bound magnetic dipoles (monopole---anti--monopole bound
states) from the vacuum while the density of unpaired monopoles (which are
responsible for the confinement of electric charges) stays unchanged. This
result is not obvious from the beginning since another option is possible: the
external fields could destroy the monopole bound states enforcing confining
properties of the medium. This is not the case.

At zero external fields the magnetic dipole states are more elongated in
the temperature direction. The external magnetic field which is parallel to
the temperature direction makes this elongation stronger. However, an
external electric field turns the dipoles  to the corresponding spatial
direction. The effects of the external field on the medium are stronger in
the deconfinement phase in which both electric and magnetic external fields
are not screened.

We have also shown that the external electric field influences the Polyakov
loop classically (or, in other words, on tree level). This leads to a
vanishing Polyakov loop and, in certain cases, to a non--vanishing
``effective string tension''~\eq{eff:sigma} depending on the external field
(being in deconfinement at zero field). However, this behavior of the
Polyakov loop does not indicate a restoration of confinement for certain
external field fluxes. At special flux values (for which the internal electric
field is quantized according to eq.~\eq{cond:N:E}) both Polyakov loop
expectation value and $\sigma_{\mathrm{eff}}$ coincide with the values at zero
external field.

The string tension (correctly defined from the correlation function of
Polyakov plane--plane correlators parallel to the external electric field
in spatial direction) is not influenced by the external electric field and
coincides with the zero--field value.

The tree level effects on Polyakov loops and Polyakov loop correlation
functions are absent for external magnetic fields pointing in the time--like
direction.
 
The dynamics of the Abelian system is different from the behavior of the
$(3+1)D$ non--Abelian gauge theory. The authors of Ref.~\cite{Cea} have found
that the external magnetic field increases the deconfinement temperature
contrary to our results in $(2+1)D$ compact Abelian gauge theory. The reason
of this difference may lie in the different behavior of the monopoles in the
Abelian and non--Abelian gauge theories. The investigation of the monopole
properties in the non--Abelian gauge theory is underway~\cite{CIS2001c}.

\section*{Acknowledgements}

M. N. Ch. acknowledges a support of S\"ach\-sis\-ches Staats\-ministerium
f\"ur Kunst und Wissenschaft, grant 4-7531.50-04-0361-01/16 and kind
hospitality of NTZ and the Institute of Theoretical Physics of Leipzig
University.  Work of M. N. Ch. was partially supported by grants RFBR
99-01230a, RFBR 01-02-17456, INTAS 00-00111 and CRDF award RP1-2103.
E.-M. I. acknowledges the support by the Graduiertenkolleg
Quantenfeldtheorie for a working visit to Leipzig.


\end{document}